\documentclass[a4paper,11pt]{article}
\usepackage{graphicx}
\usepackage{subfigure}
\usepackage{latexsym}
\usepackage[misc,geometry]{ifsym}
\usepackage{amssymb}
\usepackage{amsthm}
\usepackage{amsmath}
\DeclareMathOperator\arctanh{arctanh}
\usepackage{hyperref}
\usepackage{vmargin}
\setpapersize{A4}
\setmarginsrb{15mm}{15mm}{15mm}{15mm}{0mm}{05mm}{0mm}{10mm}
\usepackage{fancyhdr}
\usepackage{epstopdf}
\usepackage{color}
\newtheorem{proposition}{Proposition}
\newtheorem{corollary}{Corollary}

\newcommand{\expect}[1]{{\mathbb{E}}}
\newcommand{\Rea}[1]{{\mathcal{R}}}

\begin{document}

\title{\textbf{Transboundary Pollution Externalities:\\ Think Globally, Act Locally?}}

\author{Davide La Torre\footnote{SKEMA Business School - Universit$\acute{e}$ C$\hat{o}$te d'Azur, Sophia Antipolis Campus, France. Contact: \href{mailto:davide.latorre@skema.edu}{davide.latorre@skema.edu}.}
\and{Danilo Liuzzi\footnote{University of Cagliari, Department of Economics and Business, Cagliari, Italy. Contact: \href{mailto:danilo.liuzzi@unica.it}{danilo.liuzzi@unica.it}.}}
\and{Simone Marsiglio\footnote{University of Pisa, Department of Economics and Management, via Cosimo Ridolfi 10, Pisa 56124, Italy. Contact: \href{mailto:simone.marsiglio@unipi.it}{simone.marsiglio@unipi.it}.}}
}
\date{}

\maketitle

\begin{abstract}
We analyze the implications of transboundary pollution externalities on environmental policymaking in a spatial and finite time horizon setting. We focus on a simple regional optimal pollution control problem in order to compare the global and local solutions in which, respectively, the transboundary externality is and is not taken into account in the determination of the optimal policy by individual local policymakers. We show that the local solution is suboptimal and as such a global approach to environmental problems is effectively needed. Our conclusions hold true in different frameworks, including situations in which the spatial domain is either bounded or unbounded, and situations in which macroeconomic-environmental feedback effects are taken into account. We also show that if every local economy implements an environmental policy stringent enough, then the global average level of pollution will fall. If this is the case, over the long run the entire global economy will be able to achieve a completely pollution-free status. \medskip

\end{abstract}

\section{Introduction}

After decades of debates it has finally grown a shared consensus on the fact that anthropogenetic activities, and in particular economic activities, are an important determinant of environmental problems and climate change (Oreskes, 2004, IPCC, 2014). Policymakers need thus to critically intervene to reduce the accumulation of polluting emissions in the atmosphere in order to ensure that economic development is effectively addressed towards a sustainable path. However, understanding how to determine the optimal size of these policy interventions is not simple at all, especially because the stock of pollution in specific locations is strongly affected by the level of emissions generated in other locations as well, a phenomenon referred to as transboundary externality (Ansuategi and Perrings, 2000; Ansuategi, 2003). In order to ensure that such externalities are effectively accounted for in the design of environmental policy it has often been suggested that local policymakers should adopt a global perspective and collaborate with one another. This argument has been summarized in the popular motto \emph{``think globally, act locally''}\footnote{The motto ``think globally, act locally'' seems to have origins in the pioneering work of the Scottish urban planner Geedes (1915) which, even if not explicitly containing the phrase, clearly discusses the underlying argument. 
} (or \emph{``think global, act local''}). Despite the fact that the benefits of such a collaborative approach to policymaking are quite clear, understanding how to effectively implement collaboration is not simple at all, since quantifying the size of such transboundary externalities and what individual policymakers should do is all but trivial. The goal of this paper is to partly contribute to this debate by formally analyzing the desirability of a think globally, act locally approach and the type of intervention that local policymakers should opt for in order to effectively implement it.

Specifically, we analyze a pollution control problem in presence of transboundary externalities. On the one hand, several works have discussed different types of pollution control problems (Bawa, 1975; Forster, 1972, 1975; Keeler et al., 1973; van der Ploeg and Withagen, 1991; Athanassoglou and Xepapadeas, 2012; Saltari and Travaglini, 2014; La Torre et al., 2017), but to the best of our knowledge, none has ever considered the effects of transboundary externalities. On the other hand, several works have introduced transboundary externalities in the form of spatial spillovers in the contexts of capital accumulation (Brito, 2004; Camacho and Zou, 2004; Camacho et al., 2008; Boucekkine et al., 2009; 2013a, 2013b) and environmental problems (Brock and Xepapadeas, 2008, 2010; Camacho and P\'erez--Barahona, 2015; La Torre et al., 2015, 2019a, 2019b), but none in a pollution control framework. We therefore try to bridge these two branches of the economics literature by analyzing a finite horizon framework in which the social planner tries to minimize the social costs of pollution (La Torre et al., 2017) by accounting for the fact that polluting emissions, independently of where they are originated, naturally spread in the atmosphere affecting the pollution stock of every location (La Torre et al., 2015). Such a setting allows us to compare the ``global'' and ``local'' solutions in which, respectively, the transboundary externality is and is not taken into account in the determination of the optimal environmental policy by single local policymakers. This approach gives rise to a simple ``regional optimal control problem'', which is a specific type of spatially-structured optimal control problem aiming to understand whether the local solution is effectively optimal also globally (Lions, 1973, 1988). This class of problem was born as an application of the think globally, act locally motto in mathematics, thus our paper partly relates to this literature as well (see Anita and Capasso, 2018, for a recent survey of applications in epidemiology). To the best of our knowledge, ours is the first paper bringing a regional optimal control approach in economics.

The analysis of our simple regional optimal control problem, and its peculiar quadratic-linear formulation, allows us to analytically derive two interesting conclusions. First, we show that whenever some heterogeneity in the initial spatial distribution of pollution exists (which seems to be the most relevant scenario from a real-world perspective), the local solution is suboptimal and as such a global approach to environmental problems is desirable, consistently with the think globally, act locally motto. Second, we quantify the amount of collaboration that is needed at local level in order to achieve globally desirable goals. Specifically, we show that if every local economy implements an environmental policy stringent enough (and we determine what stringent enough exactly means), then the global average pollution level will fall. If this is the case, then over the long run the entire global economy will be able to achieve a completely pollution-free status. These two results jointly suggest that from a normative perspective it may well be possible to determine how global collaboration could be implemented locally in order to deal with common pollution problems. To the best of our knowledge such a neat and clear characterization of these issues from an economic point of view has never been provided before. We also show that our main conclusions do not depend on the peculiarity of our model but they rather extend to more general and complicated frameworks.

This paper proceeds as follows. Section \ref{sec:mod} discusses environmental policy in a local setting, where there are no transboundary externalities. This represents a benchmark for our following analysis, and it basically consists of an a-spatial pollution control problem over finite horizon, in which the stock of pollution is entirely determined by local economic conditions. We derive the optimal policy which characterizes the environmental tax in a setting in which local policymakers independently determine their level of intervention. Section \ref{sec:mod1} extends our baseline model in order to allow for a spatial dimension in which transboundary externalities occur as a result of spatial pollution diffusion over a bounded spatial domain. We derive the optimal policy which characterizes how the environmental tax is determined at a global level as a coordinated decision of local policymakers. We show that, in the most interesting situations in which some initial spatial heterogeneity exists, the localized approach to environmental policy gives rise to suboptimal solutions, suggesting that coordination across local policymakers is essential to deal with environmental problems. We also quantify the minimal level of intervention needed at local level allowing the global average level of pollution to decrease over time, and we show that if such a minimal level of intervention is implemented in every local economy then over the long run the entire global economy will be able to achieve a completely pollution-free status. The specific linear-quadratic formulation of our model allows us to derive the explicit solution for the spatio-temporal dynamic path of the environmental tax and the pollution level, which we illustrate through a numerical example to clearly visualize the extent to which the local and global solutions may differ. Section \ref{sec:ext} presents some extensions of our baseline model showing that our main results straightforwardly apply also in more general frameworks, including traditional macroeconomic-environmental settings with economy-environment feedbacks. Section \ref{sec:mod2} presents a further extension of our baseline model in which the spatial domain is unbounded showing that also in such a setting our main conclusions apply. Also in such a framework we derive the explicit solution for the spatio-temporal dynamic path of the environmental tax and the pollution level.
Section \ref{sec:conc} as usual concludes and suggests directions for future research. 
The proofs of most of our results are presented in appendix \ref{sec:app}.

\section{The A-Spatial Model} \label{sec:mod}

We consider a deterministic version of the finite time horizon pollution control problem recently presented in La Torre et al. (2017). Agents consume entirely their disposable income, $c_t =(1- \tau_t )y_t$, where $c_t$ is consumption, $y_t$ income and $\tau_t\in(0,1)$ the tax rate. The final consumption good, $y_t$, is produced through a linear production function, $y_t=ak_t$, where $a>0$ is a technological parameter and $k_t$ capital. Capital grows at a constant exogenous rate (normalized to unity), and productive activities generate pollution, $p_t$. The tax revenue is entirely allocated to environmental policy aiming to reduce pollution accumulation, and one unit of output devoted to environmental preservation reduces one unit of pollution. Pollution dynamics is summarized by the following equation $\dot{p}_t=\left[\eta(1-\tau_t)-\delta \right]p_t$, where $\eta>0$ is the rate at which output growth generates emissions and $\delta>0$ the natural pollution decay rate. In this setting, the policy instrument $\tau_t$ represents the environmental tax aiming to manage the economic-environmental trade off. The social planner wishes to minimize the social cost of pollution by choosing the optimal level of the tax rate. The social cost function, $\mathcal{C}$, is the weighted sum of two terms: the expected discounted ($\rho>0$ is the discount factor) sum of instantaneous social losses depending on both environmental and economic factors, and the discounted environmental damage associated with the level of pollution remaining at the end of the planning horizon, $T$. Both the loss function $c(p_t,\tau_t)$ and the damage function $d(p_T)$ are assumed to take a quadratic form as follows: $c(p_t,\tau_t)=\frac{p_t^2(1+\tau_t^2)}{2}$ and $d(p_T)=\frac{p_T^2}{2}$, respectively. The weight of the social losses and the environmental damage are given by $\theta\in[0,1]$ and $1-\theta$, respectively, such that $\frac{1-\theta}{\theta}$ represents the relative weight of the environmental damage in terms of the social losses.

Therefore, the social planner needs to choose the level of the environmental tax in order to minimize the social cost, given the evolution of pollution and its initial condition. The planner's optimal control problem reads as follows
:
\begin{eqnarray}
\min_{\tau_t}& &\mathcal{C}=\int_{0}^{T} \frac{p_t^2(1+\tau_t^2)}{2}e^{-\rho t}dt+\frac{1-\theta}{\theta}\frac{p_T^2}{2}e^{-\rho T} \label{eq:C0}\\
s.t.& &\dot{p}_t=\left[\eta(1-\tau_t)-\delta \right]p_t \label{eq:p00}\\
& &p_0>0\ \mbox{given}, \label{eq:ic00}
\end{eqnarray}
Note that the objective function in the above problem reflects sustainability considerations related to intertemporal equity (Chichilnisky et al., 1995; Chichilnisky, 1997; Colapinto et al., 2017). In particular, it is consistent with the so-called Chichilnisky's criterion which proposes to consider a weighted average between the discounted sum of instantaneous costs and the long run cost associated with pollution in order to make sure that future generations' wellbeing is effectively taken into account in the determination of the current optimal policy (Chichilnisky, 1997). The larger $1-\theta$ (i.e., the lower $\theta$) the larger the weight attached to future generations, suggesting, as we shall clarify later, that environmental policy will tend to be stricter in order to allow future generations to live in a cleaner environment; thus $1-\theta$ represents the degree of sustainability concern (La Torre et al., 2017).

In order to simplify the above problem, we can define $u_t = p_t \tau_t$ which allows us to obtain a linear-quadratic model's formulation which can be explicitly solved in closed form. Through this variable change, the above problem can be rewritten as follows:
\begin{eqnarray}
\min_{u_t}& &\mathcal{C}=\int_{0}^{T} \frac{p_t^2 + u_t^2}{2}e^{-\rho t}dt+{1-\theta\over \theta}\frac{p_T^2}{2}e^{-\rho T} \label{eq:C01}\\
s.t.& &\dot{p}_t=\left(\eta-\delta \right)p_t - \eta u_t\label{eq:p001}\\
& &p_0>0\ \mbox{given}, \label{eq:ic001}
\end{eqnarray}
The current value Hamiltonian function, $\mathcal{H}\big(p_t,u_t,\lambda_t\big)$, read as:
\begin{eqnarray*}
	\mathcal{H}&=& \frac{p_t^2 + u_t^2}{2} + \lambda_t \left[(\eta-\delta)p_t - \eta u_t\right], \nonumber
\end{eqnarray*}
where $\lambda_t$ is the costate variable. The FOCs for a minimum are given by the following expressions:
\begin{eqnarray}
\dot{\lambda}_t &=& \rho \lambda_t - p_t - (\eta-\delta)\lambda_t \label{lambdaeq}\\
u_t &=& \eta \lambda_t.
\end{eqnarray}
Substituting the latter expression in the former allows us to obtain the following system of differential equations for the state and control variables:
\begin{eqnarray}
\dot{u}_t &=& (\rho- \eta +\delta) u_t - \eta p_t  \label{ueq}\\
\dot{p}_t&=&\left(\eta-\delta \right)p_t - \eta u_t, \label{peq}
\end{eqnarray}
which, jointly with the terminal condition $u_{T}= \eta{1-\theta\over \theta} p_T$, completely characterize the optimal solution of our a-spatial control problem. By defining the following vector $z_t$ and matrix $\Theta$:
\begin{eqnarray*}
z_t= \left[
\begin{array}{c}
           p_t \\
           u_t \\
\end{array}
\right], \ \ \ \ \ \Theta= \left[
  \begin{array}{cc}
    \eta-\delta  & - \eta \\
    - \eta & \rho- \eta +\delta \\
  \end{array}
\right]
\end{eqnarray*}
the system (\ref{ueq}) and (\ref{peq}) can be rewritten as as follows:
\begin{eqnarray*}
\dot{z}_t = \Theta z_t
\end{eqnarray*}
whose solution is given by:
\begin{eqnarray*}
z_t =  C e^{\Theta t},
\end{eqnarray*}
where $C=[C_1, C_2]^T$. Since the two eigenvalues are both real and distinct, it is not complicated to calculate the exponential term $e^{\Theta t}$ as follows:
\begin{eqnarray}
 e^{\Theta t} =  \left[ \begin {array}{cc} \frac{1}{2}\,{e}^{\frac{1}{2}\,\rho\,t} \left( \cosh
 \left( \frac{1}{2}\,\xi\,t \right) -{\frac { \left(2(\delta-\eta)+\rho
 		\right) \sinh \left( \frac{1}{2}\,\xi\,t \right) }{\xi}} \right) &-{\frac {{e
 		}^{\frac{1}{2}\,\rho\,t}\sinh \left(\frac{1}{2}\,\xi\,t \right) \eta}{\xi}}
 \\ \noalign{\medskip}-{\frac {{e}^{\frac{1}{2}\,\rho\,t}\sinh \left( \frac{1}{2}\,\xi
 		\,t \right) \eta}{\xi}}&\frac{1}{2}\,{e}^{\frac{1}{2}\,\rho\,t} \left( \cosh \left( \frac{1}{2}\,\xi\,t \right) +{\frac { \left( 2(\delta-\eta)+\rho \right)
 		\sinh \left( \frac{1}{2}\,\xi\,t \right) }{\xi}} \right) \end {array} \right]
\end{eqnarray}
where $\xi=\sqrt{[2(\delta-\eta)+\rho]^2+4\eta^2}$. The explicit solution of the system above is therefore given by the following expressions:
\begin{eqnarray}
u_t^* &=& {\frac {\eta\, \left(\left( 1-\theta \right) \xi-  \left\{  \left( 1-\theta \right)  \left[ 2(\delta-\eta)+\rho \right] -2\,\theta \right\} \tanh \left[\frac{1}{2}\,\xi\,
		\left( T-t \right)  \right] \right) {
			p_0}}{{{\rm e}^{\frac{1}{2}\,\rho\,T}} \left\{ \theta\,\xi+\theta\, \left[ 2(\delta-\eta)+\rho \right]
		+2\,{\eta}^{2} \left( 1-\theta \right) \tanh \left(\frac{1}{2}\,\xi\,T
		\right)  \right\} }} \label{optu}\\
p_t^* &=& {\frac { \left\{ \theta\, \left[ 2(\delta-\eta)+\rho \right] +2\,{
			\eta}^{2} \left( 1-\theta \right) \tanh \left[ \frac{1}{2}\,\xi\, \left( T-t
		\right)  \right] +\theta\,\xi \right\} {p_0}}{{{\rm e}^{\frac{1}{2}\,\rho\,
				T}} \left\{ \theta\,\xi+\theta\, \left[ 2(\delta-\eta)+\rho \right]
		+2\,{\eta}^{2} \left( 1-\theta \right) \tanh \left(\frac{1}{2}\,\xi\,T
		\right)  \right\} }} \label{optp},
\end{eqnarray}
from which it follows that the optimal environmental tax rate, $\tau_t=\frac{u_t}{p_t}$, read as follows:
\begin{eqnarray}
\tau_t^*=\frac {\eta\, \left(\left( 1-\theta \right) \xi-  \left\{  \left( 1-\theta \right)  \left[ 2(\delta-\eta)+\rho \right] -2\,\theta \right\} \tanh \left[\frac{1}{2}\,\xi\,
		\left( T-t \right)  \right] \right)}{
		 \theta\, \left[ 2(\delta-\eta)+\rho \right] +2\,{
			\eta}^{2} \left( 1-\theta \right) \tanh \left[ \frac{1}{2}\,\xi\, \left( T-t
		\right)  \right] +\theta\,\xi } \label{opttau}
\end{eqnarray}
This expression can be rewritten as in La Torre et al. (2017) as follows:
\begin{eqnarray}
\tau_t&=&\frac{1}{2\eta} \left\{  2(\eta-\delta) -\rho + \sqrt{[2(\eta-\delta)-\rho]^2+4\eta^2} \tanh \left[ \frac{\sqrt{[2(\eta-\delta)-\rho]^2+4\eta^2}(T-t)}{2} + \right.\right.\nonumber\\
& &\left.\left. + \arctanh \left( \frac{2(1-\theta)\eta^2-2(\eta-\delta)\theta+\rho \theta}{\theta \sqrt{[2(\eta-\delta)-\rho]^2+4\eta^2}} \right) \right] \right\}
\end{eqnarray}
where $\tanh(z)=\frac{e^z-e^{-z}}{e^z+e^{-z}}$ and $\arctanh(z)=\frac{\log(1+z)-log(1-z)}{2}$, with $ -1<z<1$, are the hyperbolic tangent function and its inverse, respectively.
From the (\ref{opttau}) we can note that the optimal environmental tax is time-varying and intuitively increases with the degree of sustainability concern, and similarly also the optimal pollution level changes over time to reflect the environmental tax dynamics. Moreover, the optimal environmental tax is completely independent of the initial pollution level $p_0$, suggesting that the only determinants of the optimal policy are environmental ($\eta$ and $\delta$) and economic ($\rho$, $\theta$, $T$) parameters. Note that in the determination of the above optimal policy the social planner takes into account only the specific characteristics of the local economy, meaning that the pair $(\tau_t^*,p_t^*)$ in (\ref{opttau}) and (\ref{optp}) characterizes the local solution in which local policymakers independently determine their level of policy intervention.

\section{The Spatial Model: Bounded Spatial Domain}  \label{sec:mod1}

We now introduce a spatial dimension in the above problem to allow pollution to diffuse across space. We assume that the economy develops along a linear city (Hotelling, 1929), where all activities take place, and in particular pollution, even if generated in a specific location, diffuses across the whole economy (La Torre et al., 2015). We denote with $\tau_{x,t}$ and $p_{x,t}$ the tax rate and the pollution stock in the position $x$ at date $t$, in a compact interval $[x_a,x_b]\subset \mathbb{R}$. We also assume that the economy is isolated, in the sense that there is no flow of pollution that can cross the boundary in the normal direction and the component on the boundary is purely tangential. Mathematically this is described by the so-called Neumann's conditions, which state that the normal derivatives at $x=\{x_a, x_b\}$ are null: $\frac{\partial \tau_{x,t}}{\partial x}=\frac{\partial p_{x,t}}{\partial x}=0$. Other constraints, such us Dirichlet or Robin boundary conditions, might be imposed as well, however these conditions will somehow force the pollution level to assume a specific value at the boundary and thus to follow a specific path over time, which do not seem to fit the purpose of this paper which rather aims at exploring the pollution dynamics in a closed economy without external impositions. In this framework, any position $x$ may be interpreted as a specific local economy while the entire spatial domain as the global economy; such a possibility to distinguish between local and global economies allows us to compare the local and global solutions of the pollution control problem. Modeling the spatial domain as a compact interval implies that we are considering an isolated global economy (i.e., an island country located far away from other countries) in which pollution is entirely determined by the behavior of the different local economies which compose it (i.e., the subnational unities within the country). Despite this may seem a very restrictive assumption, as we shall see in section \ref{sec:mod2}, our main results will not depend on the boundedness of the spatial domain. In this setting, the spatial control model can be summarized by the following problem:
\begin{eqnarray}
\min_{\tau_{x,t}}& &\mathcal{C}=\int_{0}^{T}\int_{x_a}^{x_b} \frac{p_{x,t}^2(1+\tau_{x,t}^2)}{2}e^{-\rho t}dxdt+\frac{1-\theta}{\theta}\int_{x_a}^{x_b}\frac{p_{x,T}^2}{2}e^{-\rho T}dx \label{eq:C1}\\
s.t.& &\frac{\partial p_{x,t}}{\partial t} = d\frac{\partial^2 p_{x,t}}{\partial x^2}+ \left[\eta(1-\tau_{x,t})-\delta \right]p_{x,t} \label{eq:p1}\\
& &\frac{\partial \tau_{x,t}}{\partial x}=\frac{\partial p_{x,t}}{\partial x}=0, x\in\{x_a,x_b\}, \label{eq:no} \\
& &p_{x,0}>0\ \mbox{given} \label{eq:ic1}
\end{eqnarray}
With respect to the a-spatial control problem discussed in the previous section, two important differences arise. (i) The social planner wishes to minimize the social costs of pollution within the global economy (i.e., over the entire spatial domain), thus the optimal policy determined in this framework characterizes the global solution which policymakers collaborating with one another would agreed upon, or which a global intergovernmental policymaker would determine and impose to local policymakers. (ii) The dynamics of pollution is described by a partial differential equation (PDE) which characterizes its spatio-temporal evolution, and the parameter $d\geq0$ represents the diffusion coefficient which measures the speed at which the pollution stock spreads across space. Therefore, the term $d\frac{\partial^2 p_{x,t}}{\partial x^2}$ in (\ref{eq:p1}) introduces some transboundary pollution externalities since the increase in the stock of pollution in each local economy $x$ is affected by the stock of pollution in other local economies $x'\neq x$ as well. These two substantive differences with respect to the a-spatial model imply that the environmental outcomes in different local economies are all interrelated and fully accounted for in the determination of the global optimal policy; this intuitively suggests, as we shall prove later, that, generally speaking, a global approach to environmental policy is superior to a local approach.

As in our previous a-spatial analysis, we simplify the problem by defining a new control variable $u_{x,t}= p_{x,t} \tau_{x,t}$. Under this variable change, our problem reads as follows:
\begin{eqnarray}
\min_{\tau_{x,t}}& &\mathcal{C}=\int_{0}^{T}\int_{x_a}^{x_b} \frac{p_{x,t}^2+u_{x,t}^2}{2}e^{-\rho t}dxdt+\frac{1-\theta}{\theta}\int_{x_a}^{x_b}\frac{p_{x,T}^2}{2}e^{-\rho T}dx \label{eq:C11}\\
s.t.& &\frac{\partial p_{x,t}}{\partial t} = d\frac{\partial^2 p_{x,t}}{\partial x^2}+ \left(\eta -\delta \right)p_{x,t} - \eta u_{x,t} \label{eq:p11}\\
& &\frac{\partial u_{x,t}}{\partial x}=\frac{\partial p_{x,t}}{\partial x}=0, x\in\{x_a,x_b\}, \label{eq:no1} \\
& &p_{x,0}>0\ \mbox{given}. \label{eq:ic11}
\end{eqnarray}


Before deriving the global solution determined by solving the problem (\ref{eq:C11}) -- (\ref{eq:ic11}), it may be useful to discuss the local solution. In the local solution, local policymakers do not internalize the transboundary pollution externality and thus the environmental tax is determined as in the previous section (\ref{opttau}); however, differently from the previous section in which we have abstracted entirely from spatial considerations, because of such a transboundary externality, the level of pollution in each local economy strictly depends on the level of pollution in other local economies as well. Therefore, the local solution is given by the pair ($\overline{u}_{x,t}$, $\overline{p}_{x,t}$) where:
\begin{eqnarray}
	\overline{u}_{x,t}=\tau^*_t \overline{p}_{x,t}\label{optuloc}
\end{eqnarray}
and $\overline{p}_{x,t}$ is the solution of the following partial differential equation:
\begin{eqnarray}
	\frac{\partial \overline{p}_{x,t}}{\partial t} = d\frac{\partial^2 \overline{p}_{x,t}}{\partial x^2}+ \left(\eta -\delta \right)\overline{p}_{x,t} - \eta u_{x,t},
\label{optploc}
\end{eqnarray}
and $\tau^*_t $ is given in (\ref{opttau}). Therefore, in the local solution the environmental tax is set without considering any spatial interaction between local economies but pollution is transboundary because of spatial diffusion.

Moving now to the global solution, we approach the spatial optimal control problem (\ref{eq:C11}) -- (\ref{eq:ic11}) by following a variational method (Troltzsch, 2010; Boucekkine et al., 2013a). The generalized current value Hamiltonian function, $\mathcal{H}(p_{x,t},u_{x,t}, \lambda_{x,t})$, reads as follows:
\begin{eqnarray}
	\mathcal{H}&=& \frac{p_{x,t}^2+u_{x,t}^2}{2} + \lambda_{x,t} \left[d\frac{\partial^2
		p_{x,t}}{\partial x^2} +(\eta-\delta)p_{x,t} - \eta u_{x,t}\right] \nonumber
\end{eqnarray}
where $\lambda_{x,t}$ is the costate variable. The FOCs for a minimum are given by the following expressions:
\begin{eqnarray}
\frac{\partial \lambda_{x,t}}{\partial t}&=&\rho
\lambda_{x,t}-d\frac{\partial^2 \lambda_{x,t}}{\partial x^2}
- \, p_{x,t} - (\eta-\delta)\lambda_{x,t}  \label{lambdap}\\
u_{x,t}&=&\eta \lambda_{x,t}\label{tau}
\end{eqnarray}

Rearranging this last expression in terms of $\lambda_{x,t}$ and substituting this into (\ref{eq:p11}) and (\ref{lambdap}) we obtain the following system of PDEs:
\begin{eqnarray}
\frac{\partial p_{x,t}}{\partial t}&=&d\frac{\partial^2 p_{x,t}}{\partial x^2} + \left(\eta -\delta \right)p_{x,t} - \eta u_{x,t}   \label{optimal1} \\
\frac{\partial u_{x,t}}{\partial t}&=&\rho u_{x,t}-d\frac{\partial^2 u_{x,t}}{\partial x^2}- \eta p_{x,t} - (\eta-\delta) u_{x,t} ,\label{optimal2}
\end{eqnarray}
which, jointly with the following boundary conditions:
\begin{eqnarray}
	&&p_{x,0}=p_{0}(x) \label{optimal3} \\
	&& u_{x,T}= \eta \, \frac{1-\theta}{\theta} p_{x,T} \label{optimal4} \\
	&&\frac{\partial p_{x_a,t}}{\partial x}=\frac{\partial p_{x_b,t}}{\partial x}=0\ \ \  \forall t\in [0,T] \label{optimal5}\\
	&&\frac{\partial \tau_{x_a,t}}{\partial x}=\frac{\partial \tau_{x_b,t}}{\partial x}=0\ \ \  \forall t\in [0,T], \label{optimal6}
\end{eqnarray}
completely characterize the optimal solution of our spatial control problem. The analysis of the optimality conditions above allows us to derive some interesting results about the eventual differences between the local and global approaches to policymaking. These are summarized in the following Propositions \ref{prop:loc} and \ref{prop:glob}.

\begin{proposition} \label{prop:loc}
Assume that $p_{0}(x) = p_0$ along with $u_{x,t}=\overline{u}_{x,t}$ and $p_{x,t}=\overline{p}_{x,t}$ as in (\ref{optuloc}) and (\ref{optploc}), respectively.
Then the pair $(\overline{u}_{x,t},\overline{p}_{x,t})$ represents the optimal solution of the control problem (\ref{eq:C11}) -- (\ref{eq:ic11}). 
\end{proposition}

\begin{proposition}\label{prop:glob}
Assume that $p_{0}(x) \neq p_0$ along with $u_{x,t}=\overline{u}_{x,t}$ and $p_{x,t}=\overline{p}_{x,t}$ as in (\ref{optuloc}) and (\ref{optploc}).
Then the pair $(\overline{u}_{x,t},\overline{p}_{x,t})$ does not represent the optimal solution of the control problem (\ref{eq:C11}) -- (\ref{eq:ic11}). 
\end{proposition}

Proposition \ref{prop:loc} states that in a framework in which there is no spatial heterogeneity in the initial pollution distribution, then the local approach to policymaking is optimal also from a global perspective, since the global solution coincides with the local solution discussed in the previous section. Proposition \ref{prop:glob} states that, whenever there exists some spatial heterogeneity in the initial pollution distribution, the local approach to policymaking is suboptimal from a global perspective, since the local solution does not solve the spatial control problem above. These results are quite intuitive: in the absence of heterogeneity transboundary pollution externalities do not play any role, and thus local policymakers can safely determine their optimal intervention levels without taking into account what is happening in the surrounding local economies. In the presence of heterogeneity transboundary pollution externalities become critical and therefore determining local policies without accounting for them will lead to suboptimal results globally. Clearly, from a real world perspective, the existence of initial spatial heterogeneity is the most interesting and realistic situation since different local economies are characterized by idiosyncratic economic and environmental features which have determined their economic history impacting on their specific initial pollution levels.\footnote{This is consistent with the path-dependency argument frequently discussed in the economic geography literature resulting from agglomeration and external economies (Krugman, 1991; Fujita et al., 1999; Fujita and Thisse, 2002).} In such a scenario Proposition \ref{prop:glob} provides strong support for the think globally, act locally argument.


From our above discussion, it is clear that in the most realistic situations policy coordination across local economies is essential in order to deal with environmental problems. However, understanding how to implement coordination is not simple since it is not straightforward to quantify what the intervention of single local economies should be. We will now try to shed some light on this by analyzing the patterns of the pollution and the environmental tax dynamics in a framework with heterogeneous initial spatial pollution distribution. 
It is then possible to prove the following results, summarized by Propositions \ref{prop:average}, \ref{prop:pmax} and \ref{prop:pmax1}
.

\begin{proposition} \label{prop:average}
Let $(u_{x,t},p_{x,t})\ge 0$ be the globally optimal solution of the spatial optimal control problem (\ref{optimal1}) -- (\ref{optimal6}) with initial pollution level given by $p_0(x)$, along with $u_{x,t}= p_{x,t} \tau_{x,t}$ and $\tau_{min}=\min_{(x,t)\in [x_a,x_b]\times [0,T]} \tau_{x,t}$. Define the global average of pollution at the time $t\in [0,T]$ as follows:
\begin{equation}
p^{tot}_t = \int_{x_a}^{x_b} p_{x,t} dx.
\end{equation}
If $\tau_{min}> {\eta-\delta\over \eta}$, then $p^{tot}_t$ will be non-increasing over time.
\end{proposition}


\begin{proposition} \label{prop:pmax}
Let $(u_{x,t},p_{x,t})\ge 0$ be the globally optimal solution of the spatial optimal control problem (\ref{eq:C1}) -- (\ref{eq:ic1}) with initial pollution level given by $p_0(x)$. Let
$u_{x,t}= p_{x,t} \tau_{x,t}$ and $\tau_{min}=\min_{(x,t)\in [x_a,x_b]\times [0,T]} \tau_{x,t}$; then $p_{x,t}\le e^{(\eta-\delta -\eta \tau_{min})t} h_{x,t}$ where $h_{x,t}$ is the solution of the following problem:
\begin{eqnarray}
\frac{\partial h_{x,t}}{\partial t} & = & d\frac{\partial^2 h_{x,t}}{\partial x^2} \ \ \ on \ (x_a,x_b)\times (0,T) \label{model111} \\
\frac{\partial h_{x_a,t}}{\partial x} & = & \frac{\partial h_{x_b,t}}{\partial x}=0\ \ \  \forall t\in [0,T] \label{model222} \\
h_{x,0} & = &  p_{0}(x) \ \ \ in \  (x_a,x_b) \label{model333}
\end{eqnarray}
\end{proposition}

\begin{proposition} \label{prop:pmax1}
If $\tau_{min}> {\eta-\delta\over \eta}$ then $\lim_{T\to +\infty}p_{x,T}=0$ for any $x\in [x_a,x_b]$.
\end{proposition}

Proposition \ref{prop:average} identifies a lower bound for the environmental tax in any local economy allowing to achieve a reduction in the global average pollution level. If each local economy implements at any moment in time an environmental policy stringent enough ($\tau_{x,t}>{\eta-\delta\over \eta}, \forall x, t$) then it will be possible to effectively observe a pollution reduction in the global economy. This result can be interpreted from a normative perspective to determine the minimal policy that needs to be implemented locally.  Proposition \ref{prop:pmax} states that, independently of what single local economies do, the minimum of the tax rate in different local economies at different moments in time, $\tau_{min}$, allows to determine an upper bound for the pollution stock at global level. Therefore, the lowest tax rate implemented by single local economies provides us with important information about the maximal pollution level that the global economy will need to bear. Proposition \ref{prop:pmax1} states that if the minimum of the tax is large enough (i.e., $\tau_{min}> \frac{\eta-\delta}{\eta}$) then it will be possible for the global economy to achieve a completely pollution-free status in the long run. The lowest tax rate implemented by single local economies can thus be informative also of whether the pollution problem can be effectively eliminated in the long run. These three propositions jointly allow us not only to clearly understand that some collaboration across local economies is needed, but also to quantify the minimal level of policy intervention required to achieve the desirable global goal of pollution elimination.



\subsection{An Analytical Solution}

Our results thus far have been derived by focusing on the FOCs for our optimization problem. However, because the model (\ref{eq:C11}) -- (\ref{eq:ic11}) has a linear-quadratic structure, it is possible to solve it in closed form to gain some further understanding on the difference between the local and the global solutions. This result is summarized in the next two propositions.


\begin{proposition}\label{prop:solloc}
The locally optimal pair $(\overline{p}_{x,t}, \overline{u}_{x,t})$ is given by
\begin{eqnarray}
\left[
  \begin{array}{c}
    \overline{p}_{x,t} \\
    \overline{u}_{x,t} \\
  \end{array}
\right] = \left[
    \begin{array}{c}
        e^{\int_0^t \eta -\delta - \eta \tau^*_s ds} \left[ {A_0\over 2} + \sum_{n\ge 1} A_n \cos\left(2 n \pi \left[{x-x_a\over x_b-x_a}\right]\right)\right]\\
        \tau^*_t \overline{p}_{x,t}\\
    \end{array}
\right] \label{eq:solloc}
\end{eqnarray}
where
\begin{equation*}
A_0 = {2\over x_b - x_a} \int_{x_a}^{x_b} p_0(x) dx, \ \ A_n = {2\over x_b - x_a} \int_{x_a}^{x_b} p_0(x) \cos\left(2 n \pi \left[{x-x_a\over x_b-x_a}\right]\right) dx
\end{equation*}
and $\tau_t^*$ is given by (\ref{opttau}).
\end{proposition}

\begin{proposition}\label{prop:sol}
The globally optimal pair $(p_{x,t},u_{x,t})$ is given by
\begin{eqnarray}
\left[
  \begin{array}{c}
    p_{x,t} \\
    u_{x,t} \\
  \end{array}
\right] = e^{\Theta t}
\left[
  \begin{array}{c}
    {A_0\over 2} + \sum_{n\ge 1} A_n e^{-d\left({2 n\pi\over x_b-x_a}\right)^2 t} \cos\left(2 n \pi \left[{x-x_a\over x_b-x_a}\right]\right) \\
    {B_0\over 2} + \sum_{n\ge 1} B_n e^{-d\left({2 n\pi\over x_b-x_a}\right)^2 (T-t)} \cos\left(2 n \pi \left[{x-x_a\over x_b-x_a}\right]\right) \\
  \end{array}
\right]\label{eq:sol}
\end{eqnarray}
where
\begin{eqnarray}
 e^{\Theta t} =  \left[ \begin {array}{cc} \frac{1}{2}\,{e}^{\frac{\rho t}{2}} \left( \cosh
 \left( \frac{\xi t}{2} \right) -{\frac { \left(2(\delta-\eta)+\rho
 		\right) \sinh \left( \frac{\xi t}{2} \right) }{\xi}} \right) &-{\frac {{e
 		}^{\frac{\rho t}{2}}\sinh \left(\frac{\xi t}{2} \right) \eta}{\xi}}
 \\ \noalign{\medskip}-{\frac {{e}^{\frac{\rho t}{2}}\sinh \left( \frac{\xi t}{2} \right) \eta}{\xi}}&\frac{1}{2}\,{e}^{\frac{\rho t}{2}} \left( \cosh \left( \frac{\xi t}{2} \right) +{\frac { \left( 2(\delta-\eta)+\rho \right)
 		\sinh \left( \frac{\xi t}{2} \right) }{\xi}} \right) \end {array} \right]
\end{eqnarray}
and
$$
A_0 = {2\over x_b - x_a} \int_{x_a}^{x_b} p_0(x) dx \ \ \ \ A_n = {2\over x_b - x_a} \int_{x_a}^{x_b} p_0(x) \cos\left(2 n \pi \left[{x-x_a\over x_b-x_a}\right]\right) dx
$$
$$
B_0 = \left[\frac{\theta e^{\Theta T}_{21}  - \eta (1-\theta) e^{\Theta T}_{11}}{\eta (1-\theta) e^{\Theta T}_{12} - \theta e^{\Theta T}_{22}}\right]A_0 \ \ \ \ B_n = \left[\frac{\theta e^{\Theta T}_{21}  - \eta (1-\theta) e^{\Theta T}_{11}}{\eta (1-\theta) e^{\Theta T}_{12} - \theta e^{\Theta T}_{22}}\right]A_n e^{-d\left({2 n\pi\over x_b-x_a}\right)^2 T}
$$
\end{proposition}

Propositions \ref{prop:solloc} and \ref{prop:sol} determine explicitly the spatio-temporal dynamic path of the pollution level, $p_{x,t}$, and the environmental tax, $\tau_{x,t}=\frac{u_{x,t}}{p_{x,t}}$, in local and global settings, respectively. However, the expressions in (\ref{eq:solloc}) and (\ref{eq:sol}) are particularly cumbersome and thus it is not possible to perform some comparative statics exercises in order to understand how they depend on the different economic and environmental parameters. Nevertheless, they allow to explicitly verify Proposition \ref{prop:loc}. Indeed, as shown in the following corollary, in the absence of spatial heterogeneity in the initial distribution of pollution, the above dynamic paths representing the global solution of our transboundary pollution control problem boil down to those found earlier in the local solution.


\begin{corollary} \label{coroll}
Suppose $p_0(x) = p_0$ for all $x\in [x_a,x_b]$. Then the globally optimal pair $(p_{x,t},u_{x,t})$ is given by
\begin{eqnarray}
\left[
  \begin{array}{c}
    p_{x,t} \\
    u_{x,t} \\
  \end{array}
\right] = p_0 e^{\Theta t}
\left[
  \begin{array}{c}
    1  \\
   \left[\frac{\theta e^{\Theta T}_{21}  - \eta (1-\theta) e^{\Theta T}_{11}}{\eta (1-\theta) e^{\Theta T}_{12} - \theta e^{\Theta T}_{22}}\right] \\
  \end{array}
\right] \label{eq:sollocal}
\end{eqnarray}
where
\begin{eqnarray}
 e^{\Theta t} =  \left[ \begin {array}{cc} \frac{1}{2}\,{e}^{\frac{\rho t}{2}} \left( \cosh
 \left( \frac{\xi t}{2} \right) -{\frac { \left(2(\delta-\eta)+\rho
 		\right) \sinh \left( \frac{\xi t}{2} \right) }{\xi}} \right) &-{\frac {{e
 		}^{\frac{\rho t}{2}}\sinh \left(\frac{\xi t}{2} \right) \eta}{\xi}}
 \\ \noalign{\medskip}-{\frac {{e}^{\frac{\rho t}{2}}\sinh \left( \frac{\xi t}{2} \right) \eta}{\xi}}&\frac{1}{2}\,{e}^{\frac{\rho t}{2}} \left( \cosh \left( \frac{\xi t}{2} \right) +{\frac { \left( 2(\delta-\eta)+\rho \right)
 		\sinh \left( \frac{\xi t}{2} \right) }{\xi}} \right) \end {array} \right].
\end{eqnarray}
The globally optimal pair $(p_{x,t},u_{x,t})$ perfectly coincides with the locally optimal pair $(\overline{p}_{x,t},\overline{u}_{x,t})$.
\end{corollary}





In order to better understand the extent to which the local and global solutions may differ, we can exploit our closed-form solution from Propositions \ref{prop:solloc} and \ref{prop:sol}, along with Corollary \ref{coroll}, to visualize such an eventual difference. Therefore, we now present some numerical example in which the parameters take the same values employed by La Torre et al. (2017) in their calibration based on global $CO_2$ data under an intermediate degree of sustainability concern. Specifically, we set parameters as follows: $\eta=0.051$, $\delta=0.05$, $\rho=0.04$, $\theta=0.5$, $T=30$ and $d=0.01$.


\begin{figure}[h!]
	\begin{center}
		\includegraphics[width=80mm]{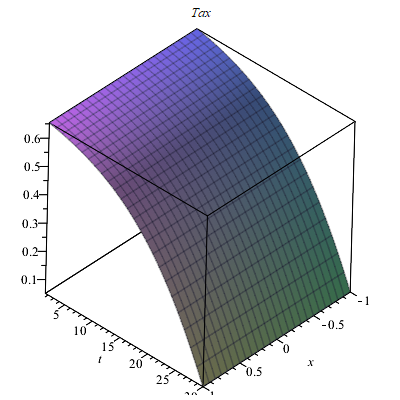}
		\includegraphics[width=80mm]{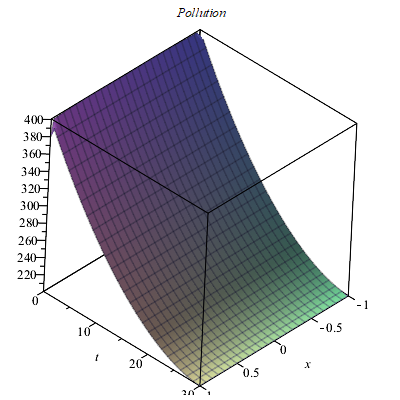}
		\caption{Spatio-temporal dynamics of the environmental tax (left) and pollution (right) with no spatial heterogeneity in the initial pollution distribution. Local and global solutions perfectly coinciding.}\label{noheter}
	\end{center}
\end{figure}

We start by illustrating the pollution and environmental tax dynamics in a framework with no spatial heterogeneity in the initial pollution distribution. Without loss of generality, we assume that $p_0(x)=p_0=400.23$, representing today's initial concentration of $CO_2$ (400.23 parts per million in year 2015). From Proposition \ref{prop:loc} (and Corollary \ref{coroll}) we know that the local and global solutions will coincide, and these identical solutions are shown in Figure \ref{noheter} where we plot the spatio-temporal evolution of the environmental tax (left panel) and pollution (right panel). Since in this setting transboundary pollution externalities do not play any role every local economy behaves exactly in the same way, and optimality implies that the tax rate initially exceeds its final level in order to effectively achieve a reduction in the stock of pollution.

\begin{figure}[h!]
	\begin{center}
		\includegraphics[width=80mm]{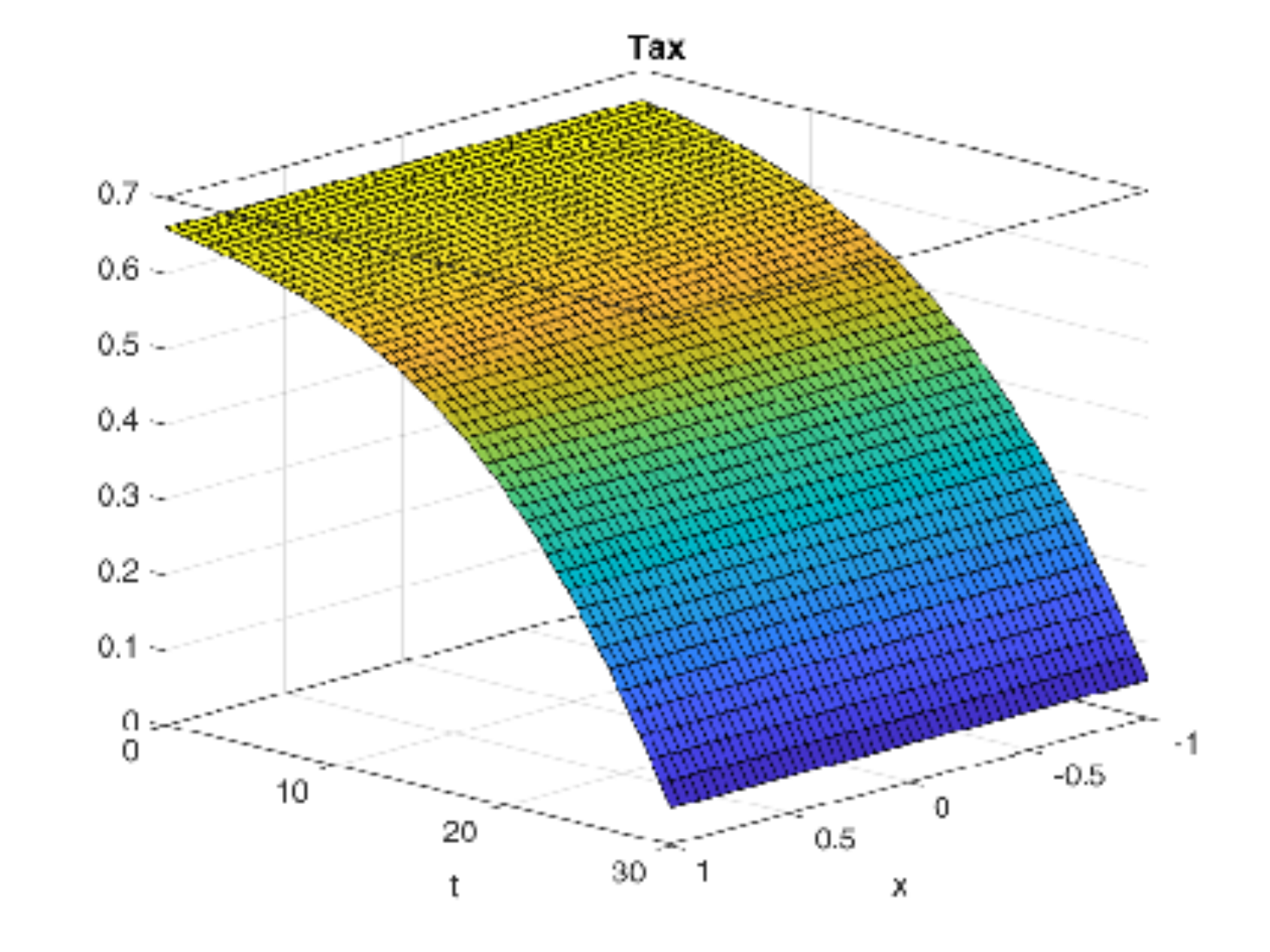}
		\includegraphics[width=80mm]{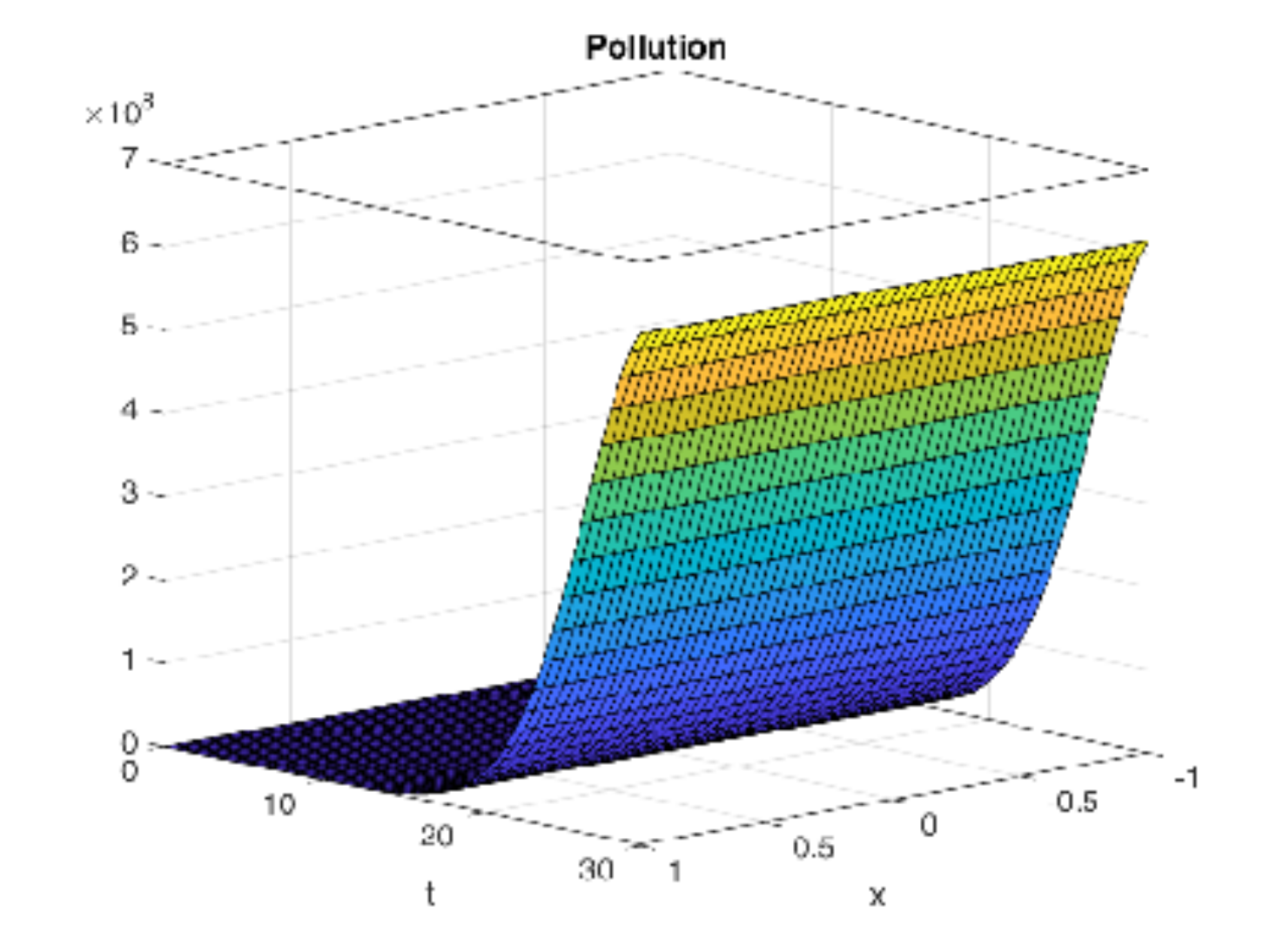}\\
        \includegraphics[width=80mm]{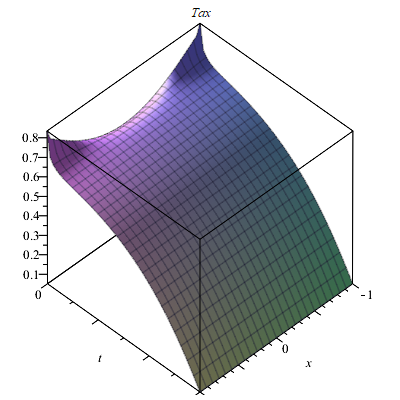}
		\includegraphics[width=80mm]{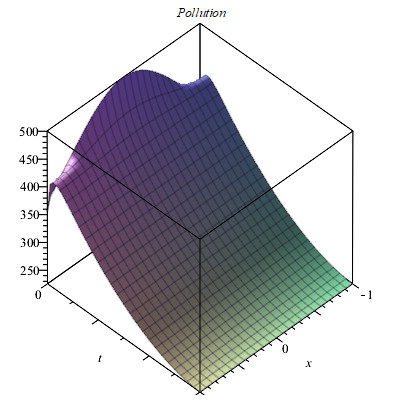}
		\caption{Spatio-temporal dynamics of the environmental tax (left) and pollution (right) with spatial heterogeneity in the initial pollution distribution. Local solution (top) and global solution (bottom) clearly not coinciding.}\label{heter}
	\end{center}
\end{figure}

We now move to the most interesting framework in which there is some spatial heterogeneity in the initial pollution distribution. In order to characterize such heterogeneity we assume that $p_0(x)=\frac{3}{4}p_0+\frac{1}{2}p_0e^{-x^2}$, suggesting that on the average in the entire global economy the concentration of $CO_2$ is roughly 400.23 parts per million, with a peak in pollution concentration in the central local economies and lower concentrations further away from the center. From Proposition \ref{prop:glob} we know that global and local solutions do not coincide and thus we will need to analyze them separately. This is shown in Figure \ref{heter} where we represent both the local solution (top panel) and the global solution (bottom panel). The local solution implies a homogeneous tax in every local economy (recall that the locally optimal tax rate in (\ref{opttau}) is independent of the initial pollution level) which however, due to the presence of transboundary pollution externalities not taken into account by local policymakers, does not allow to reduce pollution in the global economy which instead monotonically increases over time. The global solution yields a substantially different outcome: initially the environmental tax is heterogeneous and on average higher that the local tax, and such a higher tax is enough to allow pollution to monotonically fall within the global economy. The spatial heterogeneity in the optimal tax is such that the tax is lower in the central local economies where pollution concentrations are higher than in the lateral ones in which pollution concentrations are lower. This kind of counterintuitive result is due to the effect of transboundary pollution externalities which consists of homogenizing the spatial distribution of pollution: provided that the tax is high enough to ensure a reduction of pollution, in the central local economies pollution will tend to fall more rapidly than in lateral economies as a natural result of diffusion, and as such environmental policy can be less stringent in such local economies. In other words, from the global point of view of the social planner it is more effective to let diffusion do its work in the central economy rather than imposing a higher tax rate: in a sense, the social planner internalizes both the transboundary pollution externality and the physical mechanism of diffusion, allowing them to optimally interact in order to minimize the social costs. As a final remark, note that both in the local and the global solutions, at the end of the planning horizon pollution and as a result the environmental tax are spatially homogenous: this is again due to the fact that diffusion acts as a convergence mechanism which tends to smooth spatial differences out (Boucekkine et al., 2009; La Torre et al., 2015).

\section{Extensions}  \label{sec:ext}

We now consider some extensions of our baseline model in order to show that our main results related to the suboptimality of the local approach to the pollution problem still hold true even in more general settings.

\subsection{Convex Functions}

We first consider a setup in which the instantaneous loss function, the end-of-planning damage function and the pollution accumulation functions are convex. Specifically, we assume that the instantaneous loss function $c(p_{x,t},\tau_{x,t})$ is increasing and convex in both its arguments, that is $c_p>0$, $c_\tau>0$, $c_{pp}>0$ and $c_{\tau\tau}>0$, and that the end-of-planning damage function $d(p_{x,T})$ is similarly increasing and convex in pollution, that is $d>0$ and $d_{pp}>0$. We also assume that pollution accumulation is determined by a function $f(p_{x,t},\tau_{x,t})$ which is increasing in pollution and decreasing in the tax rate, $f_p>0$ and $f_\tau<0$, but convex in both its arguments, $f_{pp}\geq0$ and $f_{\tau\tau}\geq0$.


The social planner's problem in a global context reads as follows:
\begin{eqnarray}
\min_{\tau(x,t)}& &\mathcal{C}=\int_{0}^{T}\int_{x_a}^{x_b} c(p_{x,t},\tau_{x,t})e^{-\rho t}dxdt+\frac{1-\theta}{\theta}\int_{x_a}^{x_b}d(p_{x,T})e^{-\rho T}dx \\
s.t.& &\frac{\partial p_{x,t}}{\partial t} = d\frac{\partial^2 p_{x,t}}{\partial x^2}+ f(p_{x,t},\tau_{x,t})\\
& &\frac{\partial \tau(x,t)}{\partial x}=\frac{\partial p(x,t)}{\partial x}=0, x\in\{x_a,x_b\}, \\
& &p_{x,0}>0\ \mbox{given},
\end{eqnarray}
from which the following FOCs follow:
\begin{eqnarray}
c_\tau + \lambda f_\tau = 0 \label{focglobal1}\\
\frac{\partial\lambda}{\partial t} = \rho \lambda - c_p - \lambda f_p- d\frac{\partial^2 \lambda}{\partial x^2} \label{focglobal2}
\end{eqnarray}
In a local context, instead, the social planned determines the environmental tax by solving the following problem:
\begin{eqnarray}
\min_{\tau_t}& &\mathcal{C}=\int_{0}^{T} c(p_{t},\tau_{t})e^{-\rho t}dt+\frac{1-\theta}{\theta}d(p_T)e^{-\rho T} \\
s.t.& &\dot{p}_t= f(p_{t},\tau_{t}) \\
& &p_0>0\ \mbox{given},
\end{eqnarray}
from which the following FOCs follow:
\begin{eqnarray}
c_\tau + \lambda f_\tau = 0 \label{foclocal1}\\
\dot{\lambda} = \rho \lambda - c_p - \lambda f_p \label{foclocal2}
\end{eqnarray}
which determines in a completely a-spatial manner the tax, $\tau_t^*$, which in turn drives the spatio-temporal evolution of pollution as follows:
\begin{eqnarray}
\frac{\partial \overline{p}_{x,t}}{\partial t} = d\frac{\partial^2 \overline{p}_{x,t}}{\partial x^2}+ f(\overline{p}_{x,t},\tau_{t}^*).
\end{eqnarray}
By comparing (\ref{focglobal1}) - (\ref{focglobal2}) with (\ref{foclocal1})- (\ref{foclocal2}) , it is straightforward to conclude that the local solution is also a global solution if and only if $\frac{\partial^2 \lambda}{\partial x^2}=0$, that is the case whenever the initial pollution distribution is homogeneous but not whenever this is heterogeneous.
This suggests that Propositions \ref{prop:loc} and \ref{prop:glob} still hold true independently of the specific functional forms assumed in the analysis. 


\subsection{Capital Accumulation}

We now consider a setting in which there is capital accumulation and so optimal saving. This framework brings our analysis into a traditional macroeconomic context and as such we need to modify the objective function in order to represent social welfare. Social welfare is the sum of two terms: the infinite discounted sum of utilities and the end-of-planning horizon utility. The finite-time utility function $u(c_{x,t},k_{x,t}, \tau_{x,t}, p_{x,t})$ where $c_{x,t}$ denotes consumption and $k_{x,t}$ capital, is assumed to be increasing in consumption and capital but decreasing in the tax rate and pollution, and to be concave in all of its arguments. The end-of-planning horizon utility $v(k_{x,T},p_{x,T})$ is increasing in capital and decreasing in pollution and concave in both the arguments. The accumulation of capital is determined by a function $g(c_{x,t},k_{x,t}, \tau_{x,t}, p_{x,t})$ which is increasing in capital, decreasing in the remaining arguments, and concave in all of them. The accumulation of pollution function $f(k_{x,t}, \tau_{x,t}, p_{x,t})$, other than the properties earlier discussed, is increasing and concave in capital. Consistent with economic growth and environment literature (see Xepapadeas, 2005, for a survey), such a framework allows us to consider mutual economic-environmental feedbacks: the stock of capital determines the evolution of the pollution stock, which in turn affects the evolution of capital.


By assuming for the sake of simplicity that the diffusion parameter is the same for both capital and pollution, the social planner's problem in a global context reads as follows:
\begin{eqnarray}
\min_{\tau(x,t)}& &\mathcal{W}=\int_{0}^{T}\int_{x_a}^{x_b} u(c_{x,t},k_{x,t}, \tau_{x,t}, p_{x,t})e^{-\rho t}dxdt+\frac{1-\theta}{\theta}\int_{x_a}^{x_b}v(k_{x,T},p_{x,T})e^{-\rho T}dx \\
s.t.& &\frac{\partial p_{x,t}}{\partial t} = d\frac{\partial^2 p_{x,t}}{\partial x^2}+ f(c_{x,t},k_{x,t}, \tau_{x,t}, p_{x,t})\\
& &\frac{\partial k_{x,t}}{\partial t} = d\frac{\partial^2 k_{x,t}}{\partial x^2}+ g(c_{x,t},k_{x,t}, \tau_{x,t}, p_{x,t})\\
& &\frac{\partial \tau(x,t)}{\partial x}=\frac{\partial p(x,t)}{\partial x}= \frac{\partial c(x,t)}{\partial x}=\frac{\partial k(x,t)}{\partial x}0, x\in\{x_a,x_b\}, \\
& &p_{x,0}>0,\ k_{x,0}>0\ \mbox{given},
\end{eqnarray}
from which the following FOCs follow:
\begin{eqnarray}
u_c+\lambda g_c =0 \label{focglobal3}\\
u_\tau + \lambda g_\tau + \xi f_\tau= 0 \label{focglobal4}\\
\frac{\partial\lambda}{\partial t}= \rho \lambda - u_k - \lambda g_k - \xi f_k- d_k\frac{\partial^2 \lambda}{\partial x^2} \label{focglobal5}\\
\frac{\partial\xi}{\partial t} = \rho \xi - u_p - \lambda g_p - \xi f_p - d\frac{\partial^2 \xi}{\partial x^2} \label{focglobal6}
\end{eqnarray}
In a local context, instead, the social planned determines the environmental tax and the level of consumption by solving the following problem:
\begin{eqnarray}
\max_{c_t,\tau_t}& &\mathcal{W}=\int_{0}^{T} u(c_{t},k_{t}, \tau_{t}, p_{t})e^{-\rho t}dt+\frac{1-\theta}{\theta}v(k_T, p_T)e^{-\rho T} \\
s.t.& &\dot{p}_t= f(k_{t}, \tau_{t}, p_{t}) \\
& &\dot{k}_t = g(c_{t},k_{t}, \tau_{t}, p_{t})\\
& &p_0>0,\ k_0>0\ \mbox{given},
\end{eqnarray}
from which the following FOCs follow:
\begin{eqnarray}
u_c+\lambda g_c =0 \label{foclocal3}\\
u_\tau + \lambda g_\tau + \xi f_\tau= 0 \label{foclocal4}\\
\dot{\lambda} = \rho \lambda - u_k - \lambda g_k - \xi f_k \label{foclocal5}\\
\dot{\xi} = \rho \xi - u_p - \lambda g_p - \xi f_p, \label{foclocal6}
\end{eqnarray}
and this determines in a completely a-spatial manner the tax, $\tau_t^*$, and consumption, $c_t^*$, which in turn jointly drive the spatio-temporal evolution of pollution and capital as follows:
\begin{eqnarray}
\frac{\partial \overline{p}_{x,t}}{\partial t} &=& d\frac{\partial^2 \overline{p}_{x,t}}{\partial x^2}+ f(c_{t}^*,\overline{k}_{x,t}, \tau_{t}^*, \overline{p}_{x,t})\\
\frac{\partial \overline{k}_{x,t}}{\partial t} &=& d\frac{\partial^2 \overline{k}_{x,t}}{\partial x^2}+ g(c_{t}^*,\overline{k}_{x,t}, \tau_{t}^*, \overline{p}_{x,t})
\end{eqnarray}
By comparing (\ref{focglobal3}) - (\ref{focglobal6}) with (\ref{foclocal3}) - (\ref{foclocal6}), it is straightforward to conclude that the local solution is also a global solution if and only if $\frac{\partial^2\lambda}{\partial x^2}=\frac{\partial^2\xi}{\partial x^2}=0$, that is the case whenever the initial pollution and capital distributions are homogeneous, but not whenever they are heterogeneous.
This suggests that Propositions \ref{prop:loc} and \ref{prop:glob} still hold true independently of the fact that we introduce a richer macroeconomic framework with mutual economic-environmental feedback effects in the analysis. 



\section{The Spatial Model: Unbounded Spatial Domain}  \label{sec:mod2}

We now consider a further extension of our baseline model, in which the spatial domain is no longer bounded but rather unbounded. Different from what assumed earlier, now we assume that $x\in \mathbb{R}$ and thus there are no natural borders for the global economy. Modeling the spatial domain as the whole real line implies that we are considering an integrated global economy (i.e., the world economy, in which countries all interconnected one another) in which pollution, even if generated in locations very apart one another, depends on the behavior of all the different locations which compose it (i.e., the national countries within the world economy). In our baseline setup with a bounded spatial domain the behavior of pollution at the borders plays a critical role, and in particular the Neumann conditions guarantee that the endogenous patterns emerging from the optimizing choices of the social planner are not induced by setting the pollution level at some arbitrary value at the borders. In this sense, our extension to a unbounded spatial domain, in which natural borders do not exist, removes this potential problem and allows us to consider the optimally determined pollution dynamics without imposing any further restriction. In this setting the spatial control problem reads as follows:
\begin{eqnarray}
\min_{u_{x,t}}& &\mathcal{C}=\int_{0}^{T}\int_{-\infty}^{+\infty} \frac{p_{x,t}^2+u_{x,t}^2}{2}e^{-\rho t}dxdt+\frac{1-\theta}{\theta}\int_{-\infty}^{+\infty}\frac{p_{x,T}^2}{2}e^{-\rho T}dx \label{ubeq:C12}\\
s.t.& &\frac{\partial p_{x,t}}{\partial t} = d\frac{\partial^2 p_{x,t}}{\partial x^2}+ \left(\eta -\delta \right)p_{x,t} - \eta u_{x,t} \ \ \ on \ (-\infty,+\infty)\times (0,T) \label{ubeq:p12}\\
& &p_{x,0}>0\ \ \ \ in \  (-\infty,+\infty) \label{eq:ic02}
\end{eqnarray}
Most of the calculations presented in section \ref{sec:mod1} apply in this case as well, and it is possible to show that the following system of PDEs:
\begin{eqnarray}
\frac{\partial p_{x,t}}{\partial t}&=&d\frac{\partial^2 p_{x,t}}{\partial x^2} + \left(\eta -\delta \right)p_{x,t} - \eta u_{x,t}   \label{optimal11} \\
\frac{\partial u_{x,t}}{\partial t}&=&\rho u_{x,t}-d\frac{\partial^2 u_{x,t}}{\partial x^2}- \eta p_{x,t} - (\eta-\delta) u_{x,t} ,\label{optimal21}
\end{eqnarray}
jointly with the following boundary conditions:
\begin{eqnarray}
	&&p_{x,0}=p_{0}(x) \label{optimal31} \\
	&& u_{x,T}= \eta \, \frac{1-\theta}{\theta} p_{x,T} \label{optimal41}
\end{eqnarray}
characterize the optimal solution of our spatial control problem. It is straightforward to note that the presence of $d\frac{\partial^2 u_{x,t}}{\partial x^2}$ places eventually a wedge between the local and the global solution. Therefore, exactly the same results as in Propositions \ref{prop:loc} and \ref{prop:glob} hold: in the case in which the initial pollution distribution is homogeneous the local solution coincides with the global one, while in the case in which the initial pollution distribution is heterogeneous the local and the global solutions differ with the local solution being suboptimal. This suggests that our conclusions regarding the desirability of a global approach to environmental problems, consistently with the think globally, act locally motto, are independent of the specific assumptions on the structure of the spatial domain.

Moreover, by applying the same arguments employed in Propositions \ref{prop:pmax} and \ref{prop:pmax1}, it is possible to prove the following result.

\begin{proposition} \label{prop:pmax2}
Let $(u_{x,t},p_{x,t})\ge 0$ be the globally optimal solution of the spatial optimal control problem (\ref{ubeq:C12}) -- (\ref{eq:ic02}) with initial pollution level given by $p_0(x)$. Let $u_{x,t}= p_{x,t} \tau_{x,t}$, and $\tau_{min}=\min_{(x,t)\in (-\infty,+\infty)\times [0,T]} \tau_{x,t}$; then $p_{x,t}\le e^{(\eta-\delta -\eta \tau_{min})t} h_{x,t}$ where $h_{x,t}$ is the solution of the following problem:
\begin{eqnarray}
\frac{\partial h_{x,t}}{\partial t} & = & d\frac{\partial^2 h_{x,t}}{\partial x^2} \ \ \ on \ (-\infty,+\infty)\times (0,T) \label{model111} \\
h_{x,0} & = &  p_{0}(x) \ \ \ in \  (-\infty,+\infty) \label{model333}
\end{eqnarray}
where $h_{x,t}$ is then the well-known classical solution of the heat equation over an unbounded domain provided by means of the Green functions as follows:
$$
h_{x,t} = {1\over 2\sqrt{\pi d t}} \int_{-\infty}^{+\infty} e^{-{(x-y)^2\over 4dt}} p_0(y) dy.
$$
Moreover, if $\tau_{min}> {\eta-\delta\over \eta}$ then $\lim_{T\to +\infty}p_{x,T}=0$ for any $x\in \mathbb{R}$.
\end{proposition}

Consistent with what discussed in our baseline model, Proposition \ref{prop:pmax2} states that the minimum of the tax rate in different local economies at different moments in time, $\tau_{min}$, determines an upper bound for the pollution stock in each locations, and if such a minimum
of the tax is large enough (i.e., $\tau_{min}> {\eta-\delta\over \eta}$) then each location within the global economy will achieve a completely pollution-free status in the long run. This allows us to quantify the minimal level of policy intervention required to achieve the desirable goal of pollution elimination in the global economy and such a minimal level of collaboration across local economies perfectly coincide with that determined in our baseline model, suggesting that also our conclusion regarding the amount of collaboration needed to reduce global pollution is independent of the specific assumptions on the structure of the spatial domain.

\subsection{An Analytical Solution}

Even in the presence of an unbounded spatial domain, given the specific linear-quadratic structure of our model (\ref{ubeq:C12}) - (\ref{eq:ic02}), it is possible to solve it in closed form to gain some further understanding on the difference between the local and the global solutions. This result is summarized in the next two propositions.


\begin{proposition} \label{prop:sol1loc}
The locally optimal pair $(\overline{p}_{x,t},\overline{u}_{x,t})$ is given by
\begin{eqnarray}
\left[
  \begin{array}{c}
    \overline{p}_{x,t} \\
    \overline{u}_{x,t} \\
  \end{array}
\right] = \left[
    \begin{array}{c}
         {1\over 2\sqrt{\pi d t}} e^{\int_0^t \eta -\delta - \eta \tau^*_s ds}  \int_{-\infty}^{+\infty} e^{-{(x-y)^2\over 4d t}} p_0(y) dy\\
        \tau^*_t \overline{p}_{x,t}\\
    \end{array}
\right], \label{eq:sol1loc}
\end{eqnarray}
where $\tau_t^*$ is given by (\ref{opttau}).
\end{proposition}

\begin{proposition} \label{prop:sol1}
The globally optimal pair $(p_{x,t},u_{x,t})$ is given by
\begin{eqnarray}
\left[
  \begin{array}{c}
    p_{x,t} \\
    u_{x,t} \\
  \end{array}
\right] =
e^{\Theta t}
\left[\begin{array}{c}
{1\over 2\sqrt{\pi d t}} \int_{-\infty}^{+\infty} e^{-{(x-y)^2\over 4d t}} p_0(y) dy \\
{1\over 2\sqrt{\pi d (T-t)}} \int_{-\infty}^{+\infty} e^{-{(x-y)^2\over 4d(T-t)}} \tilde z^1_{y,T} dy \\
\end{array}\right] \label{eq:sol1}
\end{eqnarray}
where
$$
\tilde z^1_{x,T} = {1\over 2\sqrt{\pi d T}}  \left[{\theta e^{\Theta T}_{21}  - \eta (1-\theta) e^{\Theta T}_{11}\over \eta (1-\theta) e^{\Theta T}_{12} - \theta e^{\Theta T}_{22}}\right]\int_{-\infty}^{+\infty} e^{-{(x-y)^2\over 4dT}} p_0(y) dy
$$
and
\begin{eqnarray}
 e^{\Theta t} =  \left[ \begin {array}{cc} \frac{1}{2}\,{e}^{\frac{\rho t}{2}} \left( \cosh
 \left( \frac{\xi t}{2} \right) -{\frac { \left(2(\delta-\eta)+\rho
 		\right) \sinh \left( \frac{\xi t}{2} \right) }{\xi}} \right) &-{\frac {{e
 		}^{\frac{\rho t}{2}}\sinh \left(\frac{\xi t}{2} \right) \eta}{\xi}}
 \\ \noalign{\medskip}-{\frac {{e}^{\frac{\rho t}{2}}\sinh \left( \frac{\xi t}{2} \right) \eta}{\xi}}&\frac{1}{2}\,{e}^{\frac{\rho t}{2}} \left( \cosh \left( \frac{\xi t}{2} \right) +{\frac { \left( 2(\delta-\eta)+\rho \right)
 		\sinh \left( \frac{\xi t}{2} \right) }{\xi}} \right) \end {array} \right]
\end{eqnarray}
\end{proposition}

Propositions \ref{prop:sol1loc} and \ref{prop:sol1} determine explicitly the spatio-temporal dynamic path of the pollution level, $p_{x,t}$, and the environmental tax, $\tau_{x,t}=\frac{u_{x,t}}{p_{x,t}}$, in local and global settings, respectively. The same comments presented for the closed-form solution of our baseline model apply: the expressions in (\ref{eq:sol1loc}) and (\ref{eq:sol1}) are so cumbersome to prevent the possibility to understand how they depend on the different economic and environmental parameters.
Note that the closed-form expression for the dynamic path of the environmental rate and the pollution level in Propositions \ref{prop:sol1loc} and \ref{prop:sol1} can be expressed in a simpler form by interpreting it as the expected value of a random variable. More precisely, recall that if ${Z}$ is a standard normal distribution, its density is given by:
$$
f_Z(x) = {1\over \sqrt{2 \pi}} e^{-{z^2\over 2}}
$$
and the expected value of the composition $g({Z})$, where $g:\mathbb{R}\to \mathbb{R}$ is a real function, is given by
$$
\mathbb{E}[g(Z)]= \int_{-\infty}^{+\infty} g(s) f_Z(s) ds
$$
By introducing the change of variable $z = {y - x \over \sqrt{2dt}}$ and replacing $dy = \sqrt{2dt} dz$, $(\overline{p}_{x,t},\overline{u}_{x,t})$ in (\ref{eq:sol1loc}) can be rewritten as follows:
\begin{eqnarray}
\left[
  \begin{array}{c}
    \overline{p}_{x,t} \\
    \overline{u}_{x,t} \\
  \end{array}
\right] = \left[
    \begin{array}{c}
         {1\over 2\sqrt{\pi d t}} e^{\int_0^t \eta -\delta - \eta \tau^*_s ds}  \int_{-\infty}^{+\infty} e^{-{(x-y)^2\over 4d t}} p_0(y) dy\\
        \tau^*_t \overline{p}_{x,t}\\
    \end{array}
\right] = \left[
    \begin{array}{c}
         e^{\int_0^t \eta -\delta - \eta \tau^*_s ds} \mathbb{E} p_0(x+2\sqrt{d t} Z)\\
        \tau^*_t \overline{p}_{x,t}\\
    \end{array}
\right],
\end{eqnarray}
while substituting in (\ref{eq:sol1}) leads to the following:
\begin{eqnarray}
\left[
  \begin{array}{c}
    p_{x,t} \\
    u_{x,t} \\
  \end{array}
\right] =
e^{\Theta t}
\left[\begin{array}{c}
{1\over \sqrt{2 \pi}} \int_{-\infty}^{+\infty} e^{-{z^2\over 2}} p_0(x+z \sqrt{2dt}) dz\\
{1\over 2\sqrt{\pi d (T-t)}} \int_{-\infty}^{+\infty} e^{-{(x-y)^2\over 4d(T-t)}} \tilde z^1_{y,T} dy \\
\end{array}\right] =
e^{\Theta t}
\left[\begin{array}{c}
\expect(p_0(x+ \sqrt{2dt} {Z})) \\
\expect(\tilde z^1(x+ \sqrt{2d(T-t)} {Z},T)) \\
\end{array}\right]
\end{eqnarray}
where
$$
\tilde z^1_{x,T} = \tilde z^1(x,T) = \left[{\theta e^{\Theta T}_{21}  - \eta (1-\theta) e^{\Theta T}_{11}\over \eta (1-\theta) e^{\Theta T}_{12} - \theta e^{\Theta T}_{22}}\right] \expect(p_0(x+ \sqrt{2dT} {Z}))
$$

The above stochastic formulations in terms of the expected value of a rescaled normal distribution allows to express the closed-form solutions in local and global settings in a more compact form and to use a Monte-Carlo simulations to approximate the optimal paths in our following numerical example. From an economic perspective, the pollution expression states that, as expected, in the long-run the level of pollution in a single location $x$ will get more and more affected by pollution externalities over time. Despite the difficulties in interpreting the spatio-temporal dynamic path of the pollution level and the environmental tax, Propositions \ref{prop:sol1loc} and \ref{prop:sol1} allow to explicitly verify Proposition \ref{prop:loc}. Indeed, as shown in the following corollary, in the absence of spatial heterogeneity in the initial distribution of pollution, the globally optimal dynamic path of our transboundary pollution control problem boils down to the locally optimal one.

\begin{corollary} \label{coroll1}
Suppose that $p_0(x)=p_0$ for all $x\in (-\infty,+\infty)$. The globally optimal pair $(p_{x,t},u_{x,t})$ is given by
$$
\left[
  \begin{array}{c}
    p_{x,t} \\
    u_{x,t} \\
  \end{array}
\right] =
p_0 e^{\Theta t}
\left[\begin{array}{c}
1 \\
\left[{\theta e^{\Theta T}_{21}  - \eta (1-\theta) e^{\Theta T}_{11}\over \eta (1-\theta) e^{\Theta T}_{12} - \theta e^{\Theta T}_{22}}\right]\\
\end{array}\right]
$$
where
\begin{eqnarray}
 e^{\Theta t} =  \left[ \begin {array}{cc} \frac{1}{2}\,{e}^{\frac{\rho t}{2}} \left( \cosh
 \left( \frac{\xi t}{2} \right) -{\frac { \left(2(\delta-\eta)+\rho
 		\right) \sinh \left( \frac{\xi t}{2} \right) }{\xi}} \right) &-{\frac {{e
 		}^{\frac{\rho t}{2}}\sinh \left(\frac{\xi t}{2} \right) \eta}{\xi}}
 \\ \noalign{\medskip}-{\frac {{e}^{\frac{\rho t}{2}}\sinh \left( \frac{\xi t}{2} \right) \eta}{\xi}}&\frac{1}{2}\,{e}^{\frac{\rho t}{2}} \left( \cosh \left( \frac{\xi t}{2} \right) +{\frac { \left( 2(\delta-\eta)+\rho \right)
 		\sinh \left( \frac{\xi t}{2} \right) }{\xi}} \right) \end {array} \right].
\end{eqnarray}
The globally optimal pair $(p_{x,t},u_{x,t})$ perfectly coincides with the locally optimal pair $(\overline{p}_{x,t},\overline{u}_{x,t})$.
\end{corollary}

\begin{figure}[h!]
	\begin{center}
		\includegraphics[width=80mm]{tau_bounded_unbounded_local.eps}
		\includegraphics[width=80mm]{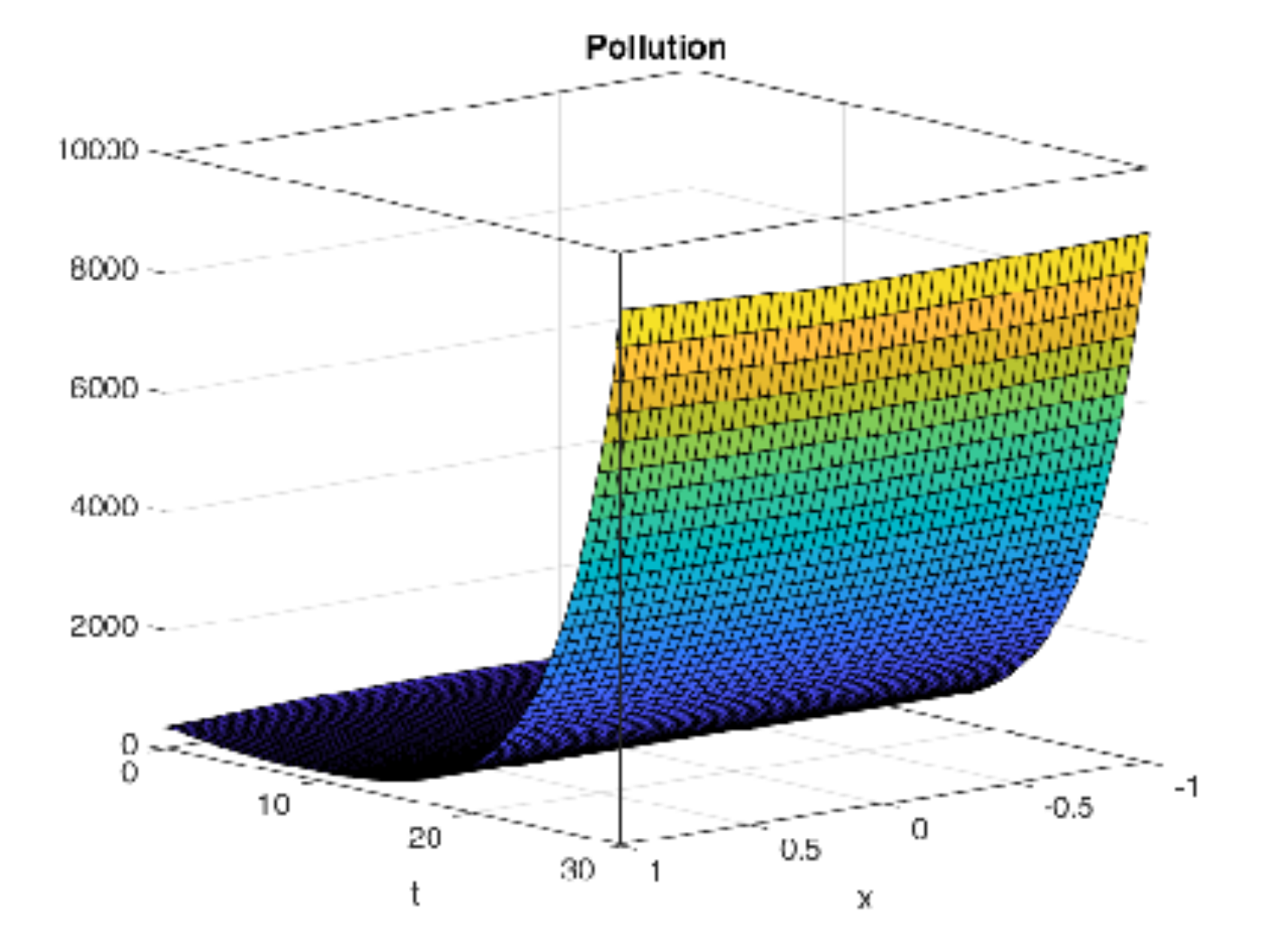}\\
        \includegraphics[width=80mm]{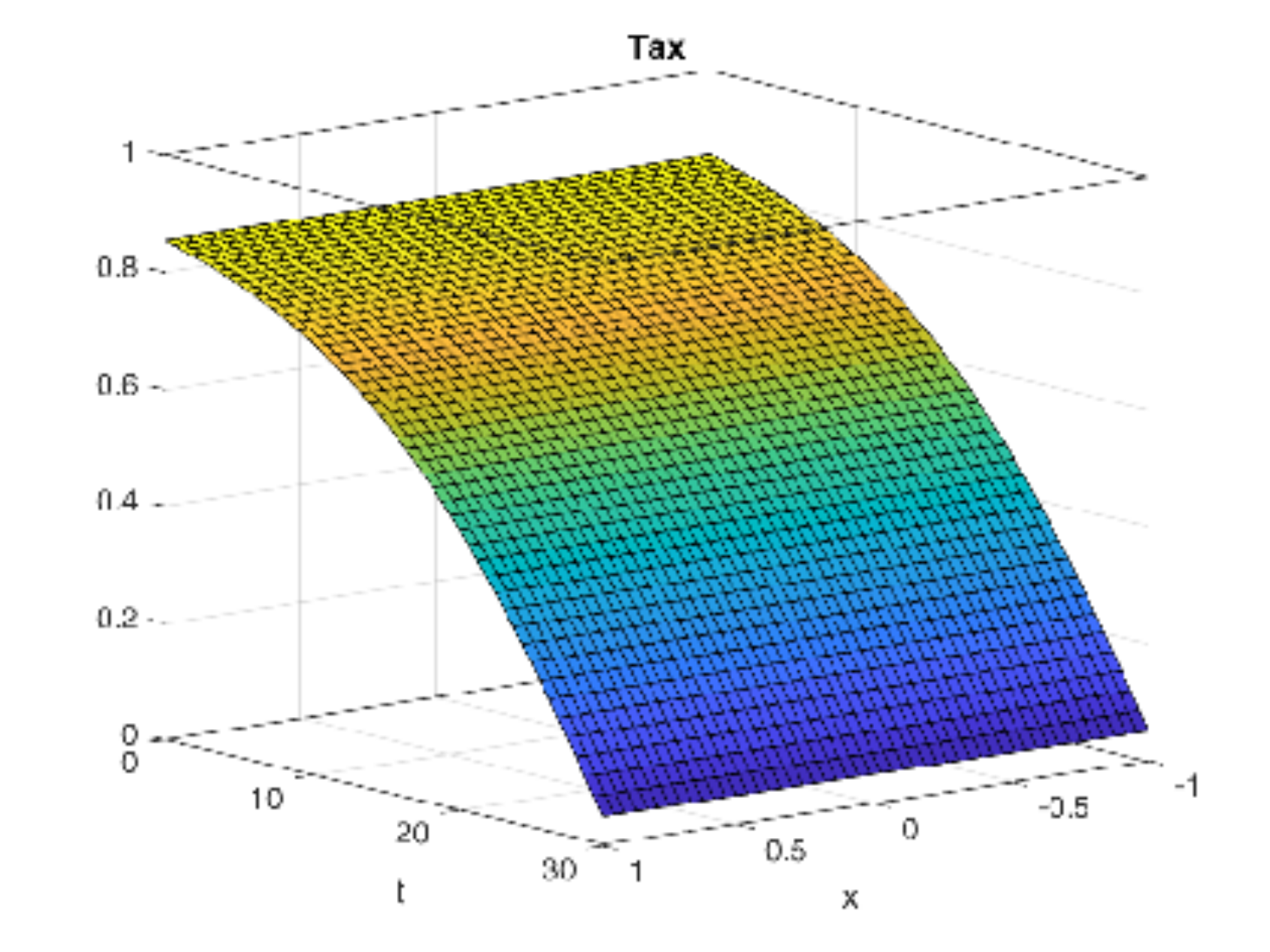}
		\includegraphics[width=80mm]{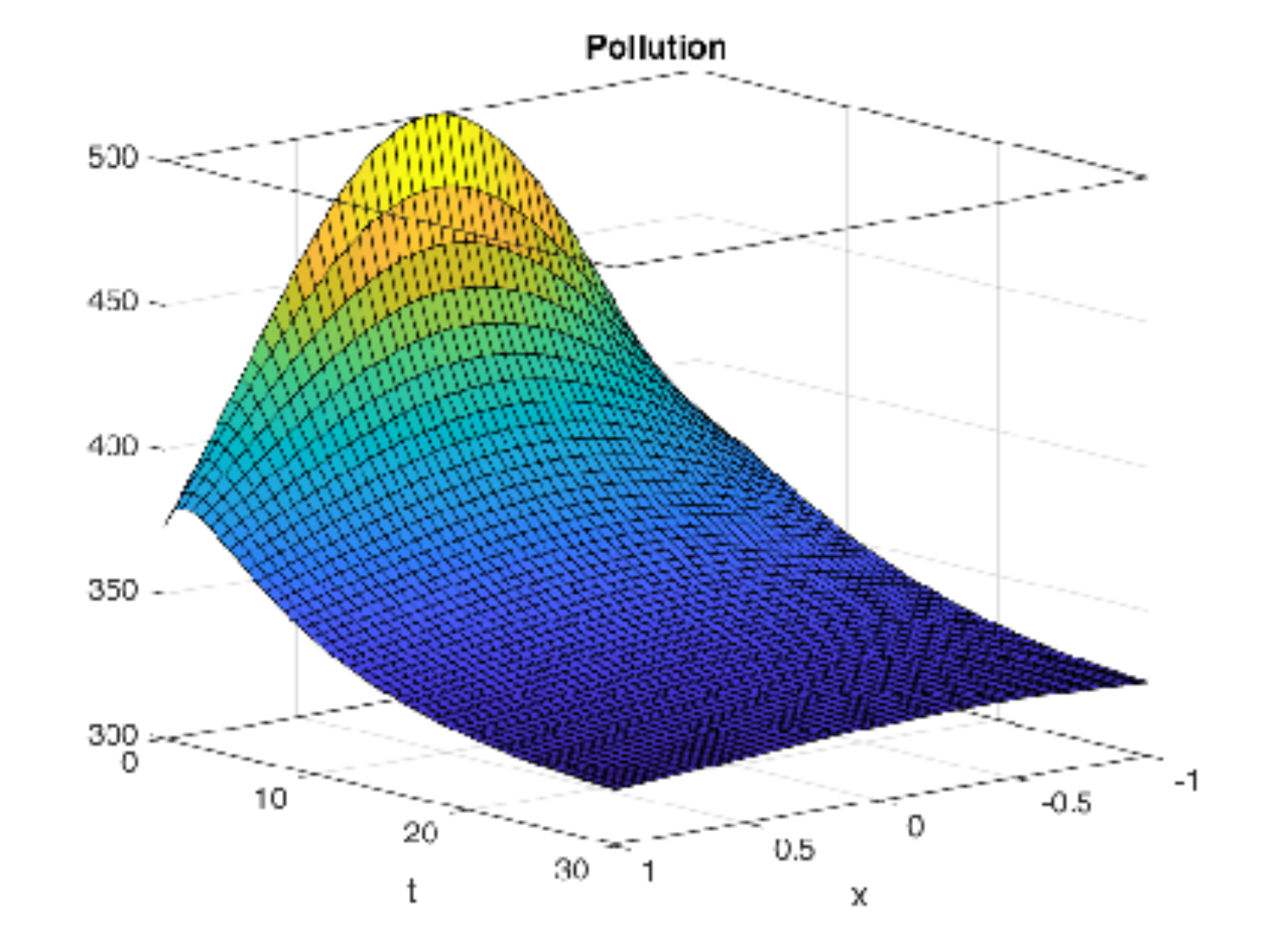}
		\caption{Spatio-temporal dynamics of the environmental tax (left) and pollution (right) with spatial heterogeneity in the initial pollution distribution. Local solution (top) and global solution (bottom) clearly not coinciding.}\label{heterunb}
	\end{center}
\end{figure}

In order to better understand the extent to which the local and global solutions may differ, we can exploit our closed-form solution from Propositions \ref{prop:sol1loc} and \ref{prop:sol1} to visualize such an eventual difference. Therefore, we now present some numerical example based on the same parameter values employed in our baseline model, by focusing only on the most interesting case in which the initial pollution distribution is heterogeneous, namely $p_0(x)=\frac{3}{4}p_0 + \frac{1}{2}p_0 e^{-x^2}$ with $x\in\mathbb{R}$ (while in our baseline model $x\in[-1,1]$). This is illustrated in Figure \ref{heterunb} where we represent both the local solution (top panel) and the global solution (bottom panel) in the spatial interval [-1, 1], that is a small subset of the entire spatial domain. Exactly as in our baseline model, in which the spatial domain is bounded, the local and global solutions yield completely different outcomes. The local solution implies a homogeneous tax in every local economy which however is not high enough to reduce pollution in the global economy. The pollution dynamics is non-monotonic as it initially decreases and then increases. Also the global solution implies a spatially homogeneous tax but this is initially higher than in the local solution, and such a higher tax is enough to yield a monotonic reduction in pollution within the global economy. The main difference with respect to the bounded spatial domain case is associated with the initial spatial structure of the environmental tax: this is heterogeneous in the bounded case and homogeneous in the unbounded case, and this is due to the role of diffusion, which per se tends to reduce pollution everywhere in the spatial interval considered and as such there is no need for the optimal tax to be space-dependent. On average the tax is higher in the unbounded than in the bounded case but nevertheless this does not lead to a lower pollution level at the end of the planning horizon, which turns out to be higher in the unbounded than in the bounded case: this is due to the initial stock of pollution which is overall larger in the unbounded that in the bounded case, since pollution concentrations are strictly positive on the whole real axis in the unbounded case while they are irrelevant outside the interval [-1,1] in the bounded case (since pollution flows stop at the border). The higher initial stock of pollution in the unbounded case requires thus a higher environmental tax on average but, despite pollution tends to smooth spatial differences out by reducing pollution concentrations in the local economies where it is relatively more abundant and increasing them in those in which it is relatively more scarse, this is not enough to offset the difference in the initial stock of pollution which remains higher in the unbounded than in the bounded case even at the end of the planning horizon.

\subsection{Extensions}

By following exactly the same arguments presented earlier in the bounded spatial domain case, it is straightforward to show that the results of Propositions \ref{prop:loc} and \ref{prop:glob} apply also in more general settings than the specific one just discussed. In particular, even in an unbounded spatial domain framework, by allowing the instantaneous loss function, the end-of-planning damage function and the pollution accumulation functions to be convex, the FOCs in the local and global cases will be given by (\ref{foclocal1})- (\ref{foclocal2}) and (\ref{focglobal1}) - (\ref{focglobal2}), respectively. By introducing capital accumulation and optimal consumption, along the lines discussed in section \ref{sec:ext}, the FOCs in the local and global cases will be given by (\ref{foclocal3}) - (\ref{foclocal6}) and (\ref{focglobal3}) - (\ref{focglobal6}), respectively. Therefore, exactly the same comments we highlighted earlier hold true, confirming that the suboptimality of the local approach to the pollution problem still applies in general settings and independently of the specific structure of the spatial domain.

\section{Conclusion} \label{sec:conc}

Environmental policy is essential in order to achieve sustainable development but understanding the optimal level of policy intervention is not simple, especially because of the presence of transboundary externalities. The popular motto think globally, act locally summarizes the view that policy coordination across individual policymakers is necessary to preserve our common environment. In this paper we try to formally analyze whether this is really the case by focusing on a pollution control problem over a finite horizon and in a spatial framework. This setting gives rise to a regional optimal control problem which allows us to compare the local and global solutions in which, respectively, the transboundary externality is and is not taken into account in the determination of the optimal policy by individual local policymakers. We show that whenever the initial spatial distribution of pollution is homogeneous then the local and the global solutions coincide, while whenever this is heterogeneous the two solutions differ meaning that the local solution is suboptimal. In this latter context, which represents the most realistic from a real world perspective, coordination across local policymakers is the best approach consistently with the think globally, act locally argument. We also quantify the minimal level of policy intervention that needs to be implemented locally in order to achieve the globally desirable goal of pollution elimination. Indeed, we show that whenever every local economy implements an environmental policy stringent enough, then the global pollution level will fall and over the long run the entire global economy will be able to achieve a completely pollution-free status. We also show that our main conclusions do not depend on the peculiarity of our model's formulation but they rather extend to more general and complicated frameworks.

To the best of our knowledge, no other paper has explicitly analyzed a regional optimal control problem in economics nor the potential differences between local and global solutions arising from an economic problem in spatial settings in a way comparable to ours. The approach has therefore been a bit simplistic and the analysis could be extended along multiple directions. Some further heterogeneity in the characteristics or the objectives of local policymakers can be introduced in order to represent the realistic situation in which some local economies (due for example to binding budget constraints) in the entire global economy cannot optimally determine their level of intervention. This further issue is left for future research.

\appendix

\section{Technical Appendix} \label{sec:app}

In this section we present all the proofs of the results discussed in the paper.

\subsection*{Proof of Proposition \ref{prop:loc}}

The proof is trivial and it follows by noticing that the pair $(\overline{u},\overline{p})$ solves equations (\ref{optimal1}) and (\ref{optimal2}), since $\frac{\partial^2 p_{x,t}}{\partial x^2}=0$, and it satisfies the boundary conditions (\ref{optimal3})-(\ref{optimal6}).

\subsection*{Proof of Proposition \ref{prop:glob}}

The proof is trivial and it follows by noticing that the pair $(\overline{u},\overline{p})$ does not solve equations (\ref{optimal1}) and (\ref{optimal2}) because
$\frac{\partial^2 p_0(x)}{\partial x^2}\neq0$.

\subsection*{Proof of Proposition \ref{prop:average}}

As $(u,p)$ is optimal, then the following inequality is true:
\begin{equation}
\frac{\partial p_{x,t}}{\partial t}\le \frac{\partial^2p_{x,t}}{\partial x^2} +(\eta-\delta - \eta \tau_{min})p_{x,t}
\end{equation}
By integrating from $x_a$ to $x_b$ we obtain:
\begin{equation}
{d\over dt} \int_{x_a}^{x_b} p_{x,t} dx \le \int_{x_a}^{x^b} \frac{\partial^2p_{x,t}}{\partial x^2} +(\eta-\delta - \eta \tau_{min}) \int_{x_a}^{x_b} p_{x,t} dx,
\end{equation}
which implies:
\begin{equation}
{d\over dt} p^{tot}_t \le (\eta-\delta -\eta \tau_{min})p^{tot}_t \le 0,
\end{equation}
and then the thesis follows.

\subsection*{Proof of Proposition \ref{prop:pmax}}

The following proposition will be instrumental to proving Proposition \ref{prop:pmax}.

\begin{proposition} (Friedman, 2008)
\label{comparison}
Let $\rho$ be smooth and suppose that:\label{prop:comp}
\begin{eqnarray}
{\partial \rho \over \partial t} - d {\partial^2 \rho \over \partial x^2} & \ge & - c \rho \ \ \ in \ (x_a,x_b)\times (0,T) \\
{\partial \rho \over \partial n} &\ge & 0, \ \ \ on \ \{x_a,x_b\}\times (0,T) \\
\rho(0, x) &\ge &  0 \ \ \ in \  (x_a,x_b)
\end{eqnarray}
where $d$ is a positive real number and $c\in\ R$. Then $\rho\ge 0$ in $(x_a,x_b)\times (0,T)$.
\end{proposition}

From the above proposition it is now straightforward to prove our thesis. Let $\bar p_{x,t}$ be a solution to the following problem:
\begin{eqnarray}
\frac{\partial p_{x,t}}{\partial t} &=& d\frac{\partial^2p_{x,t}}{\partial x^2} +(\eta-\delta-\eta \tau_{min})p_{x,t} \label{pbar1} \\
	p_{x,0} &=& p_{0}(x) \label{pbar2} \\
    \frac{\partial p_{x_a,t}}{\partial x} &=& \frac{\partial p_{x_b,t}}{\partial x}=0\ \ \  \forall t\in [0,T] \label{pbar3}
\end{eqnarray}
Let us define $\rho_{x,t} = \bar p_{x,t}-p_{x,t}$. By computing the equation for $\rho$,  we get:
\begin{eqnarray}
\frac{\partial \rho_{x,t}}{\partial t} & \ge & d\frac{\partial^2 \rho_{x,t}}{\partial x^2} + (\eta-\delta - \eta \tau_{min})\rho_{x,t} \\
	\rho_{x,t} &=& 0 \\
    \frac{\partial \rho_{x_a,t}}{\partial x} &=& \frac{\partial \rho_{x_b,t}}{\partial x}=0\ \ \  \forall t\in [0,T]
\end{eqnarray}
From the above result Proposition \ref{comparison}, we can deduce that $\rho\ge 0$ and then $p_{x,t}\le \bar p_{x,t}$.
On the other hand, by using a simple change of variable it is easy to show that if $\bar p$ solves the above problem given by (\ref{pbar1}) - (\ref{pbar3}), then $\bar p$ can be written as follows:
\begin{equation}
\bar p_{x,t} = e^{(\eta-\delta -\eta \tau_{min})t} h_{x,t}
\end{equation}
where $h_{x,t}$ is the well-known classical solution of the heat equation with Neumann boundary conditions, given by the following expression:
\begin{equation}
h_{x,t} = \sum_{n\ge 0} B_n e^{-d \left({n\pi\over x_b-x_a}\right)^2 t}\cos\left[ {n\pi(x-x_a)\over x_b-x_a}\right]
\end{equation}
where:
\begin{equation}
B_0 = {1\over x_b - x_a} \int_{x_a}^{x_b} p_{0}(x) dx
\end{equation}
and:
\begin{equation}
B_n = {2\over x_b - x_a} \int_{x_a}^{x_b} p_{0}(x) \cos\left[ {n\pi(x-x_a)\over x_b-x_a}\right] dx
\end{equation}

\subsection*{Proof of Proposition \ref{prop:pmax1}}

The proposition follows as a corollary of Proposition \ref{prop:pmax} which states that $p_{x,t}\le e^{(\eta-\delta -\eta \tau_{min})t} h_{x,t}$. If $\tau_{min}> {\eta-\delta\over \eta}$, for any $x\in [x_a,x_b]$ it is straightforward to conclude the following:
\begin{equation}
\lim_{T\to \infty} p_{x,T} \le \lim_{T\to \infty} e^{(\eta-\delta -\eta \tau_{min})T} h_{x,T} = 0.
\end{equation}

\subsection*{Proof of Proposition \ref{prop:solloc}}

To determine the closed-form local solution for the pollution level, let us plug the expression $\overline{u}_{x,t}=\tau^*_t \overline{p}_{x,t}$ into the equation for $\overline{p}$ which then boils down to
\begin{equation}
\frac{\partial \overline{p}_{x,t}}{\partial t} = d\frac{\partial^2 \overline{p}_{x,t}}{\partial x^2}+ \left(\eta -\delta \right)\overline{p}_{x,t} - \eta \tau^*_t \overline{p}_{x,t}
\end{equation}
The evolution of $\overline{p}_{x,t}$ is given by
\begin{equation}
\overline{p}_{x,t} = e^{\int_0^t \eta -\delta - \eta \tau^*_s ds} \overline{h}_{x,t}
\end{equation}
where $\overline{h}_{x,t}$ is the solution to the classical heat equation
\begin{equation}
\frac{\partial \overline{h}_{x,t}}{\partial t} = d\frac{\partial^2 \overline{h}_{x,t}}{\partial x^2}
\end{equation}
subject to the conditions:
\begin{eqnarray}
& &\frac{\partial \overline{h}_{x,t}}{\partial x}=0, x\in\{x_a,x_b\}, \label{eq:hno1} \\
& &p_{x,0}>0\ \mbox{given}. \label{eq:hic11}
\end{eqnarray}
Therefore, the solution $\overline{p}$ is given by
\begin{equation}
\overline{p}_{x,t} = e^{\int_0^t \eta -\delta - \eta \tau^*_s ds} \left[ {A_0\over 2} + \sum_{n\ge 1} A_n \cos\left(2 n \pi \left[{x-x_a\over x_b-x_a}\right]\right)\right]
\end{equation}
where
\begin{equation}
A_0 = {2\over x_b - x_a} \int_{x_a}^{x_b} p_0(x) dx, \ \ A_n = {2\over x_b - x_a} \int_{x_a}^{x_b} p_0(x) \cos\left(2 n \pi \left[{x-x_a\over x_b-x_a}\right]\right) dx
\end{equation}

\subsection*{Proof of Proposition \ref{prop:sol}}

By defining the vector-value variable $z_{x,t}$ and the matrix $\Theta$ as for the one dimensional case,
$$z_{x,t}= \left[
\begin{array}{c}
           p_{x,t} \\
           u_{x,t} \\
\end{array}
\right], \ \ \ \ \
\Theta= \left[
  \begin{array}{cc}
    \eta-\delta  & - \eta \\
    - \eta & \rho- \eta +\delta \\
  \end{array}
\right]
$$
and the diffusion matrix $D$ as follows
$$
D= \left(
     \begin{array}{cc}
       d & 0 \\
       0 & -d \\
     \end{array}
   \right)
$$
and the system can be written in a more compact form as
$$
\frac{\partial z_{x,t}}{\partial t} = D \frac{\partial^2 z_{x,t}}{\partial x^2} + \Theta z_{x,t}
$$
This is a system of two heat equations. By using the change of variable $\tilde z_{x,t} = e^{-\Theta t} z_{x,t}$
it is easy to show that the solution to this equation takes the following form:
$$
z_{x,t} = e^{\Theta t} \tilde z_{x,t}
$$
where
\begin{eqnarray}
 e^{\Theta t} =  \left[ \begin {array}{cc} \frac{1}{2}\,{e}^{\frac{\rho t}{2}} \left( \cosh
 \left( \frac{\xi t}{2} \right) -{\frac { \left(2(\delta-\eta)+\rho
 		\right) \sinh \left( \frac{\xi t}{2} \right) }{\xi}} \right) &-{\frac {{e
 		}^{\frac{\rho t}{2}}\sinh \left(\frac{\xi t}{2} \right) \eta}{\xi}}
 \\ \noalign{\medskip}-{\frac {{e}^{\frac{\rho t}{2}}\sinh \left( \frac{\xi t}{2} \right) \eta}{\xi}}&\frac{1}{2}\,{e}^{\frac{\rho t}{2}} \left( \cosh \left( \frac{\xi t}{2} \right) +{\frac { \left( 2(\delta-\eta)+\rho \right)
 		\sinh \left( \frac{\xi t}{2} \right) }{\xi}} \right) \end {array} \right]
\end{eqnarray}
and
$$
\tilde z_{x,t} =
\left[
  \begin{array}{c}
    {A_0\over 2} + \sum_{n\ge 1} A_n e^{-d\left({2 n\pi\over x_b-x_a}\right)^2 t} \cos\left(2 n \pi \left[{x-x_a\over x_b-x_a}\right]\right) \\
    {B_0\over 2} + \sum_{n\ge 1} B_n e^{-d\left({2 n\pi\over x_b-x_a}\right)^2 (T-t)} \cos\left(2 n \pi \left[{x-x_a\over x_b-x_a}\right]\right) \\
  \end{array}
\right]
$$
If we plug $t=0$ we get that $e^{\Theta 0} = I$ and then the first component of $z$ boils down to
$$
\tilde z^0_{x,0} = p_0(x) = {A_0\over 2} + \sum_{n\ge 1} A_n \cos\left(2 n \pi \left[{x-x_a\over x_b-x_a}\right]\right)
$$
which implies that $A_0$ and $A_n$ are the Fourier coefficients of $p_0$, that is
$$
A_0 = {2\over x_b - x_a} \int_{x_a}^{x_b} p_0(x) dx
$$
and
$$
A_n = {2\over x_b - x_a} \int_{x_a}^{x_b} p_0(x) \cos\left(2 n \pi \left[{x-x_a\over x_b-x_a}\right]\right) dx
$$
For the terminal condition, instead, let us plug $t=T$ into the expression of $z_{x,t}$. We get
$$
z_{x,T} = e^{\Theta T} \left[
                        \begin{array}{c}
                          \tilde z^0_{x,T} \\
                          \tilde z^1_{x,T} \\
                        \end{array}
                      \right]
= e^{\Theta T}
\left[
  \begin{array}{c}
    {A_0\over 2} + \sum_{n\ge 1} A_n e^{-d\left({2 n\pi\over x_b-x_a}\right)^2 T} \cos\left(2 n \pi \left[{x-x_a\over x_b-x_a}\right]\right) \\
    {B_0\over 2} + \sum_{n\ge 1} B_n  \cos\left(2 n \pi \left[{x-x_a\over x_b-x_a}\right]\right) \\
  \end{array}
\right]
$$
and by using the terminal condition
$$
u_{x,T}= \eta \, \frac{1-\theta}{\theta} p_{x,T}
$$
we get the system
$$
 {\theta \over \eta (1-\theta)}= \frac{e^{\Theta T}_{11} ({A_0\over 2} + \sum_{n\ge 1} A_n e^{-d\left({2 n\pi\over x_b-x_a}\right)^2 T} \cos\left(2 n \pi \left[{x-x_a\over x_b-x_a}\right]\right)) + e^{\Theta T}_{12} ({B_0\over 2} + \sum_{n\ge 1} B_n  \cos\left(2 n \pi \left[{x-x_a\over x_b-x_a}\right]\right))}{e^{\Theta T}_{21}({A_0\over 2} + \sum_{n\ge 1} A_n e^{-d\left({2 n\pi\over x_b-x_a}\right)^2 T} \cos\left(2 n \pi \left[{x-x_a\over x_b-x_a}\right]\right)) + e^{\Theta T}_{22}({B_0\over 2} + \sum_{n\ge 1} B_n  \cos\left(2 n \pi \left[{x-x_a\over x_b-x_a}\right]\right))}
$$
which can be transformed into
$$
{\theta \over \eta (1-\theta)} = \frac{e^{\Theta T}_{11} {A_0\over 2} + e^{\Theta T}_{12} {B_0\over 2} + \sum_{n\ge 1} \left(e^{\Theta T}_{11} A_n e^{-d\left({2 n\pi\over x_b-x_a}\right)^2 T} +  e^{\Theta T}_{12} B_n\right)  \cos\left(2 n \pi \left[{x-x_a\over x_b-x_a}\right]\right)}{e^{\Theta T}_{21} {A_0\over 2} + e^{\Theta T}_{22} {B_0\over 2} + \sum_{n\ge 1} \left(e^{\Theta T}_{21}  A_n e^{-d\left({2 n\pi\over x_b-x_a}\right)^2 T} + e^{\Theta T}_{22} B_n \right) \cos\left(2 n \pi \left[{x-x_a\over x_b-x_a}\right]\right)}
$$
and then
$$
 \theta \left\{e^{\Theta T}_{21} {A_0\over 2} + e^{\Theta T}_{22} {B_0\over 2} + \sum_{n\ge 1} \left(e^{\Theta T}_{21}  A_n e^{-d\left({2 n\pi\over x_b-x_a}\right)^2 T} + e^{\Theta T}_{22} B_n \right) \cos\left(2 n \pi \left[{x-x_a\over x_b-x_a}\right]\right)\right\}  =
$$
$$
 \eta (1-\theta) \left\{e^{\Theta T}_{11} {A_0\over 2} + e^{\Theta T}_{12} {B_0\over 2}+ \sum_{n\ge 1} \left(e^{\Theta T}_{11} A_n e^{-d\left({2 n\pi\over x_b-x_a}\right)^2 T} +  e^{\Theta T}_{12} B_n\right)  \cos\left(2 n \pi \left[{x-x_a\over x_b-x_a}\right]\right)\right\}
$$
and then
$$
\theta \left(e^{\Theta T}_{21} {A_0\over 2} + e^{\Theta T}_{22} {B_0\over 2} \right) = \eta (1-\theta) \left(e^{\Theta T}_{11} {A_0\over 2} + e^{\Theta T}_{12} {B_0\over 2}\right)
$$
$$
(\theta e^{\Theta T}_{21}  - \eta (1-\theta) e^{\Theta T}_{11}) {A_0\over 2} = (\eta (1-\theta) e^{\Theta T}_{12} - \theta e^{\Theta T}_{22}) {B_0\over 2}
$$
$$
B_0 = \left[\frac{\theta e^{\Theta T}_{21}  - \eta (1-\theta) e^{\Theta T}_{11}}{\eta (1-\theta) e^{\Theta T}_{12} - \theta e^{\Theta T}_{22}}\right]A_0
$$
$$
(\theta e^{\Theta T}_{21}  - \eta (1-\theta) e^{\Theta T}_{11}) A_n e^{-d\left({n\pi\over x_b-x_a}\right)^2 T} = \eta (1-\theta) e^{\Theta T}_{12} B_n - \theta e^{\Theta T}_{22} B_n
$$
$$
(\theta e^{\Theta T}_{21}  - \eta (1-\theta) e^{\Theta T}_{11}) A_n e^{-d\left({n\pi\over x_b-x_a}\right)^2 T} = (\eta (1-\theta) e^{\Theta T}_{12} - \theta e^{\Theta T}_{22}) B_n
$$
$$
B_n = \left[\frac{\theta e^{\Theta T}_{21}  - \eta (1-\theta) e^{\Theta T}_{11}}{\eta (1-\theta) e^{\Theta T}_{12} - \theta e^{\Theta T}_{22}}\right]A_n e^{-d\left({2 n\pi\over x_b-x_a}\right)^2 T}
$$

\subsection*{Proof of Proposition \ref{prop:sol1loc}}

A local solution is a pair $(\overline{u}_{x,t},\overline{p}_{x,t})$ where $\overline{u}_{x,t}=\tau^*_t \overline{p}_{x,t}$ and $\tau^*_t$ is given in (\ref{opttau}). To determine the evolution of $\overline{p}$, let us proceed as in the bounded case and plug the expression $\overline{u}_{x,t}=\tau^*_t \overline{p}_{x,t}$ into the equation for $\overline{p}$ which then boils down to
\begin{equation}
\frac{\partial \overline{p}_{x,t}}{\partial t} = d\frac{\partial^2 \overline{p}_{x,t}}{\partial x^2}+ \left(\eta -\delta \right)\overline{p}_{x,t} - \eta \tau^*_t \overline{p}_{x,t}
\end{equation}
The evolution of $\overline{p}_{x,t}$ is given by
\begin{equation}
\overline{p}_{x,t} = e^{\int_0^t \eta -\delta - \eta \tau^*_s ds} \overline{h}_{x,t}
\end{equation}
where $\overline{h}_{x,t}$ is the solution to the classical heat equation
\begin{equation}
\frac{\partial \overline{h}_{x,t}}{\partial t} = d\frac{\partial^2 \overline{h}_{x,t}}{\partial x^2}
\end{equation}
subject to the initial condition $\overline{h}_{x,0} = p_{x,0}>0$. By combining these results we get that $\overline{p}$ is given by
\begin{equation}
\overline{p}_{x,t} = {1\over 2\sqrt{\pi d t}} e^{\int_0^t \eta -\delta - \eta \tau^*_s ds}  \int_{-\infty}^{+\infty} e^{-{(x-y)^2\over 4d t}} p_0(y) dy
\end{equation}

\subsection*{Proof of Proposition \ref{prop:sol1}}

By repeating the same steps as in the baseline model, the system can be written in a more compact form as
$$
\frac{\partial z_{x,t}}{\partial t} = D \frac{\partial^2 z_{x,t}}{\partial x^2} + \Theta z_{x,t}
$$
and by introducing the change of variable $\tilde z = e^{-\Theta t} z$ it is easy to show that the solution to this equation takes the following form:
$$
z_{x,t} = e^{\Theta t} \tilde z_{x,t}.
$$
Now the first component of $\tilde z$, $\tilde z^0_{x,t}$, solves the following heat equation
$$
\frac{\partial \tilde z^0_{x,t}}{\partial t} = d\frac{\partial^2 \tilde z^0_{x,t}}{\partial x^2}
$$
with initial condition $z^0_{x,0} = p_0(x)$. The solution is then provided by means of the Green functions as follows:
$$
\tilde z^0_{x,t} = {1\over 2\sqrt{\pi d t}} \int_{-\infty}^{+\infty} e^{-{(x-y)^2\over 4d t}} p_0(y) dy
$$
The second component of $\tilde z$, $\tilde z^1_{x,t}$, satisfies the following backward heat equation
$$
\frac{\partial \tilde z^1_{x,t}}{\partial t} = - d\frac{\partial^2 \tilde z^1_{x,t}}{\partial x^2}
$$
and the solution is then provided by means of the Green functions as follows:
$$
\tilde z^1_{x,t} = {1\over 2\sqrt{\pi d(T-t)}} \int_{-\infty}^{+\infty} e^{-{(x-y)^2\over 4d(T-t)}} \tilde z^1_{y,T} dy
$$
Using the terminal condition for $u$ and $p$, we get
$$
{\theta \over \eta (1-\theta)}={p_{x,T}\over u_{x,T}} = {e^{\Theta T}_{11} \tilde z^0_{x,T} + e^{\Theta T}_{12} \tilde z^1_{x,T} \over e^{\Theta T}_{21} \tilde z^0_{x,T} + e^{\Theta T}_{22} \tilde z^1_{x,T}}
$$
which implies that
$$
\theta {e^{\Theta T}_{21} \tilde z^0_{x,T} + \theta e^{\Theta T}_{22} \tilde z^1_{x,T}} =  \eta (1-\theta) {e^{\Theta T}_{11} \tilde z^0_{x,T} + \eta (1-\theta) e^{\Theta T}_{12} \tilde z^1_{x,T}}
$$
$$
(\theta e^{\Theta T}_{21} - \eta (1-\theta) e^{\Theta T}_{11}) \tilde z^0_{x,T}  =  (\eta (1-\theta) e^{\Theta T}_{12} - \theta e^{\Theta T}_{22}) \tilde z^1_{x,T}
$$
and, finally,
$$
\tilde z^1_{x,T} = \left[{\theta e^{\Theta T}_{21}  - \eta (1-\theta) e^{\Theta T}_{11}\over \eta (1-\theta) e^{\Theta T}_{12} - \theta e^{\Theta T}_{22}}\right] \tilde z^0_{x,T}
$$

\section*{References}

\begin{enumerate}
\itemsep -0.1cm
    \item Anita, S., Capasso, V. (2018). The interplay between models and public health policies: regional control for a class of spatially structured epidemics (think globally, act locally), Mathematical Biosciences and Engineering 15, 1--20
    \item Ansuategi, A. (2003). Economic growth and transboundary pollution in Europe: an empirical analysis, Environmental and Resource Economics 26, 305--328
    \item Ansuategi, A., Perrings , C.A. (2000). Transboundary externalities in the environmental transition hypothesis, Environmental and Resource Economics 17, 353-–373
    \item Athanassoglou, S., Xepapadeas, A. (2012). Pollution control with uncertain stock dynamics: when, and how, to be precautious, Journal of Environmental Economics and Management 63, 304-–320
    \item Bawa, V.S. (1975). On optimal pollution control policies, Management Science 21, 1397--1404
    \item Boucekkine, R., Camacho, C., Zou, B. (2009). Bridging the gap between growth theory and economic geography: the spatial Ramsey model, Macroeconomic Dynamics 13, 20--45
    \item Boucekkine, R., Camacho, C., Fabbri, G. (2013a). On the optimal control of some parabolic differential equations arising in economics, Serdica Mathematical Journal 39, 331--354
    \item Boucekkine, R., Camacho, C., Fabbri, G. (2013b). Spatial dynamics and convergence: the spatial AK model, Journal of Economic Theory 148, 2719--2736
    \item Brito, P. (2004). The dynamics of growth and distribution in a spatially heterogeneous world, UECE-ISEG, Technical University of Lisbon
    \item Brock, W.A., Xepapadeas, A. (2004). Spatial analysis: development of descriptive and normative methods with applications to economic--ecological modelling, Working Papers 2004.159, Fondazione Eni Enrico Mattei
    \item Brock, W.A., Xepapadeas, A. (2008). Diffusion-induced instability and pattern formation in infinite horizon recursive optimal control, Journal of Economic Dynamics \& Control  32, 2745--2787
    \item Brock, W.A., Xepapadeas, A. (2010). Pattern formations, spatial externalities and regulation in a coupled economic--ecological system, Journal of Environmental Economics and Management 59, 149--164
    \item Camacho, C., P\'erez--Barahona, A. (2015). Land use dynamics and the environment, Journal of Economic Dynamics \& Control 52, 96--118
    \item Camacho, C., Zou, B. (2004). The spatial Solow model,  Economics Bulletin 18, 1--11
    \item Camacho, C., Zou, B., Briani, M. (2008). On the dynamics of capital  accumulation across space, European Journal of Operational Research 186 2, 451--465
    \item Chichilnisky, G., Heal, G., Beltratti, A. (1995). The green golden rule, Economics Letters 49, 174--179
    \item Chichilnisky, G. (1997). What is sustainable development?, Land Economics 73, 476--491
    \item Colapinto, C., Liuzzi, D., Marsiglio, S. (2017). Sustainability and intertemporal equity: a multicriteria approach, Annals of Operations Research 251, 271--284
    \item Forster, B.A. (1972). A note on the optimal control of pollution, Journal of Economic Theory 5, 537--539
    \item Forster, B.A. (1975).Optimal pollution control with a nonconstant exponential rate of decay, Journal of Environmental Economics and Management 2, 1--6
    \item Friedman, A. (2008). Partial differential equations of parabolic type (Dover Ed.)
    \item Fujita, M., Krugman, P., Venables, A. (1999). The spatial economy. Cities, regions and international trade (MIT Press)
    \item Fujita, M., Thisse, J.F. (2002). Economics of agglomeration (Cambridge University Press)
    \item Geddes, P. (1915). Cities in evolution (London: Williams)
    \item Hotelling, H. (1929). Stability in competition, Economic Journal 39, 41--57
    \item Intergovernmental Panel on Climate Change, IPCC (2014). Climate change 2014: synthesis report summary for policymakers (IPCC, Geneva), available at: \href{https://www.ipcc.ch/pdf/assessment-report/ar5/syr/AR5_SYR_FINAL_SPM.pdf}{https://www.ipcc.ch/pdf/assessment-report/ar5/syr/AR5\_SYR\_FINAL\_SPM.pdf}
    \item Keeler, E., Spencer, M., Zeckhauser, R. (1973). The optimal control of pollution, Journal of Economic Theory 4, 19--34
    \item Krugman, P. (1991). Increasing returns and economic geography, Journal of Political Economy 99, 483--499
    \item La Torre, D., Liuzzi, D., Marsiglio, S. (2015). Pollution diffusion and abatement activities across space and over time, Mathematical Social Sciences 78, 48--63
    \item La Torre, D., Liuzzi, D., Marsiglio, S. (2017). Pollution control under uncertainty and sustainability concern, Environmental and Resource Economics 67, 885–-903
    \item La Torre, D., Liuzzi, D., Marsiglio, S. (2019a). Population and geography do matter for sustainable development, Environment and Development Economics 24, 201-–223
    \item La Torre, D., Liuzzi, D., Marsiglio, S. (2019b). The optimal population size under pollution and migration externalities: a spatial control approach, Mathematical Modelling of Natural Phenomena 14, 104
    \item Lions, J.L. (1973).  The optimal control of distributed systems,  Russian Mathematical Surveys 28, 13--46
    \item Lions, J.L. (1988). Controlabilit\'e exacte, stabilisation et perturbation de syst\'emes distribu\'es (Masson: Paris)
    \item McAsey, M., Mou, L., Han, W. (2012). Convergence of the forward--backward sweep method in optimal control, Computational Optimization and Applications 53, 207--226
    \item Oreskes, N. (2004). The scientific consensus on climate change, Science 306, 1686
   \item Saltari, E., Travaglini, G. (2014). Pollution control under emission constraints: switching between regimes, Energy Economics, forthcoming
   \item Troltzsch, F. (2010). Optimal control of partial differential equations: theory, methods and applications (American Mathematical Society)
    \item van der Ploeg, F., Withagen, C. (1991). Pollution control and the Ramsey problem, Environmental and Resource Economics 1, 215--236
    \item Xepapadeas, A. (2005). Economic growth and the environment, in (M\"aler, K.G., Vincent, J., Eds.), ``Handbook of Environmental Economics'', vol. 3., ch. 23, 1219--1271 (Elsevier: Amsterdam, Netherlands)
\end{enumerate}

\end{document}